\input amstex
\documentstyle{amsppt}
\magnification=\magstephalf
 \addto\tenpoint{\baselineskip 15pt
  \abovedisplayskip18pt plus4.5pt minus9pt
  \belowdisplayskip\abovedisplayskip
  \abovedisplayshortskip0pt plus4.5pt
  \belowdisplayshortskip10.5pt plus4.5pt minus6pt}\tenpoint
\pagewidth{6.5truein} \pageheight{8.9truein}
\subheadskip\bigskipamount
\belowheadskip\bigskipamount
\aboveheadskip=3\bigskipamount
\catcode`\@=11
\def\output@{\shipout\vbox{%
 \ifrunheads@ \makeheadline \pagebody
       \else \pagebody \fi \makefootline 
 }%
 \advancepageno \ifnum\outputpenalty>-\@MM\else\dosupereject\fi}
\outer\def\subhead#1\endsubhead{\par\penaltyandskip@{-100}\subheadskip
  \noindent{\subheadfont@\ignorespaces#1\unskip\endgraf}\removelastskip
  \nobreak\medskip\noindent}
\outer\def\enddocument{\par
  \add@missing\endRefs
  \add@missing\endroster \add@missing\endproclaim
  \add@missing\enddefinition
  \add@missing\enddemo \add@missing\endremark \add@missing\endexample
 \ifmonograph@ 
 \else
 \vfill
 \nobreak
 \thetranslator@
 \count@\z@ \loop\ifnum\count@<\addresscount@\advance\count@\@ne
 \csname address\number\count@\endcsname
 \csname email\number\count@\endcsname
 \repeat
\fi
 \supereject\end}
\catcode`\@=\active
\CenteredTagsOnSplits
\NoBlackBoxes
\nologo
\def\today{\ifcase\month\or
 January\or February\or March\or April\or May\or June\or
 July\or August\or September\or October\or November\or December\fi
 \space\number\day, \number\year}
\define\({\left(}
\define\){\right)}
\define\Ahat{{\hat A}}

\define\CC{{\Bbb C}}

\define\Map{\operatorname{Map}}
\define\Met{\operatorname{Met}}
\define\QQ{{\Bbb Q}}
\define\RP{{\Bbb R\Bbb P}}
\define\RR{{\Bbb R}}
\define\SS{\Bbb S}
\define\Spin{\operatorname{Spin}}

\define\Tr{\operatorname{Tr}}
\define\ZZ{{\Bbb Z}}
\define\[{\left[}
\define\]{\right]}
\define\ch{\operatorname{ch}}
\define\chiup{\raise.5ex\hbox{$\chi$}}
\define\cir{S^1}

\define\dbar{{\bar\partial}}

\define\exertag #1#2{#2\ #1}

\define\ind{\operatorname{ind}}
\define\inv{^{-1}}
\define\mstrut{^{\vphantom{1*\prime y}}}
\define\protag#1 #2{#2\ #1}
\define\rank{\operatorname{rank}}
\define\res#1{\negmedspace\bigm|_{#1}}
\define\temsquare{\raise3.5pt\hbox{\boxed{ }}}

\define\theprotag#1 #2{#2~#1}

\define\xca#1{\removelastskip\medskip\noindent{\smc%
#1\unskip.}\enspace\ignorespaces }

\define\zmod#1{\ZZ/#1\ZZ}

\define\zt{\zmod2}

\NoRunningHeads 

\define\bigstrut{\vrule height12pt depth3pt width0pt} 
 
\define\tinystrut{\vrule height1pt depth3pt width0pt} 
\define\Bev{B^{\text{ev}}}
\define\Det{\operatorname{Det}}
\define\Dirac{D\hskip-.65em /}
\define\Kbar{\overline{K}\,}
\define\Kbig{\overline{K_g}\,\inv }

\define\Lt{\tilde{L}}
\define\Pf{\operatorname{Pfaff}}
\define\Sd{\Sigma ^d}
\define\TT{\Bbb{T}}
\define\Xd{X^d}
\define\abt{\alpha ,\beta ,\theta }
\define\ah{\hat\alpha }
\define\alim{\operatorname{a-lim}}
\define\bS{\partial \Sigma }
\define\bX{\partial X}
\define\bh{\hat\beta }
\define\bo{\bold{1}}
\define\bsX{\partial \scrX}
\define\curv#1{\Omega^{#1}}
\define\hol{\operatorname{hol}}
\define\id{\operatorname{id}}
\define\pf{\operatorname{pfaff}}
\define\pt{\operatorname{pt}}
\define\qz{\QQ/\ZZ}
\define\sDirac{D\hskip-.54em /}
\define\scrL{\Cal{L}}
\define\scrX{\Cal{X}}
\define\spinc{\operatorname{Spin}^c}

\input epsf

\refstyle{A}
\widestnumber\key{SSSSSSSSS}   
\document

	\topmatter
 \title\nofrills Anomalies in String Theory with D-Branes\endtitle
 \author Daniel S. Freed\\Edward Witten  \endauthor
 \thanks The first author is supported by NSF grant DMS-9626698,
 the second  by NSF grant PHY-95-13835. \endthanks
 \affil Department of Mathematics, University of Texas at Austin\\
        School of Natural Sciences, Institute for Advanced Study \endaffil
 \address Department of Mathematics, University of Texas, Austin, TX
78712\endaddress
 \email dafr\@math.utexas.edu \endemail
 \address Institute for Advanced Study, Olden Lane, Princeton NJ,
08540\endaddress
 \email witten\@sns.ias.edu \endemail
 \date July 15, 1999 \enddate
 \dedicatory For Sir Michael Atiyah \enddedicatory
	\endtopmatter

\document

 \head
 Introduction
 \endhead
 \comment
 lasteqno @  1
 \endcomment

This paper is devoted to studying global anomalies in the worldsheet path
integral of Type~II superstring theory in the presence of $D$-branes.  We
will not consider orientifolds (or Type~I superstrings) and so our string
worldsheets will be oriented Riemann surfaces, mapped to a spacetime manifold
$Y$, which is endowed with a spin structure since the model contains
fermions.  The first case to consider is that of closed worldsheets, without
boundary.  In this case, global anomalies cancel, and the worldsheet measure
is globally well-defined, given that $Y$ is spin.  This is the content of
Corollary 4.7 of the present paper; for $Y=\RR^{10}$ one has (Theorem 4.8)
the further result, related to conformal invariance, that the global anomaly
vanishes separately for left- and right-moving degrees of freedom.  By
contrast, for heterotic strings, global anomaly cancellation gives a
restriction that is not evident from the point of view of the low energy
effective field theory \cite{W2,F2}.

But we will find that Type II global worldsheet anomalies give some novel
results when $D$-branes are present.  Thus, we assume the existence of an
oriented submanifold $Q$ of spacetime on which strings can end.  We then
consider a string worldsheet consisting of an oriented surface $\Sigma$ that
is mapped to $Y$ with $\partial \Sigma$ mapped to $Q$.  On $Q$, there is a
field $A$ that is conventionally regarded as a $U(1)$ gauge field, though as
we will explain this is not the correct interpretation globally.  Another
important part of the story is the Neveu-Schwarz $B$-field, which propagates
in the bulk of spacetime.  Assume for simplicity that~$B=0$.  In this case,
the worldsheet measure contains two interesting factors which we write as
follows:
  $$ \pf(D)\cdot\exp \left(i\oint_{\partial\Sigma}A\right).  \tag{1} $$
Here $\pf(D)$ is the pfaffian (or square root of the determinant) of the
worldsheet Dirac operator $D$.  The second factor is, in the conventional
interpretation, the holonomy of $A$ around the boundary of $\Sigma$.

Our main result, \theprotag{5.6} {Theorem}, computes the anomaly in~$\pf(D)$.
Namely, $\pf(D)$~is naturally not a function, but rather a section of a line
bundle over the space of parameters---the space of maps of the worldsheet
into spacetime and worldsheet metrics.  This bundle carries a natural metric
and connection, and the anomaly is the obstruction to the existence of a
global flat section of unit norm.  We will show that this connection is flat,
but there is holonomy~$\pm 1$ determined by the second Stiefel-Whitney
class~$w_2(\nu )$ of the normal bundle~$\nu $ to the $D$-brane~$Q$ in
spacetime.\footnote{This problem exemplifies the difference between the {\it
topological\/} anomaly and the {\it geometric\/} anomaly: the topological
isomorphism class of this line bundle is determined by the third
Stiefel-Whitney class~$W_3(\nu )$, and this may vanish even if the holonomy
is nontrivial.}

This means that for~\thetag{1} to be well-defined (for~$B=0$) there must be a
compensating anomaly in the $A$-field.  If $W_3(\nu )=0$, then this is
achieved by interpreting~$A$ as a $\spinc$~connection rather than an ordinary
$U(1)$~gauge field.  When $W_3(\nu )\not= 0$ (and $B$ is topologically
trivial) the anomaly rules out~$Q$ as a possible $D$-brane in the theory.
This result from string perturbation theory matches the nonperturbative
description of $D$-brane charge as an element of $K$-theory~\cite{MM,W4}: the
normal bundle~$\nu $ must have $W_3(\nu)=0$ in order to define the charge.
If $B\not= 0$ there is another term in~\thetag{1} and a correspondingly more
complicated interpretation of the $A$-field, leading to a generalization of
the condition~$W_3(\nu) =0$ which is stated in equations~\thetag{1.12} and
\thetag{6.9}.  Also, when $B\not= 0$, $D$-brane charge takes values in a
twisted form of $K$-theory, as explained in section 5.3 of \cite{W4}.  The
net effect is always that the $Q$'s that make sense in perturbative string
theory are the ones that have Thom classes in the appropriate $K$-group.

Therefore, our major tasks are to prove the anomaly formula and to properly
interpret the $A$- and $B$-fields.  Since the results here are of direct
interest in string theory, we begin in section~1 by explaining the physical
implications of the anomaly.  Here we analyze the $B\not= 0$ case as well as
the simpler case when $B$~vanishes.  We give examples to show that the
anomaly can occur, and we also show how it relates to $D$-brane charge.
Section~2 contains a heuristic argument for the anomaly in terms perhaps more
palatable to physicists than the proofs which follow.  From a mathematical
point of view the proof, which is contained in section~5, combines ideas from
topological index theory (1960s) and geometric index theory (1980s).  The
pertinent background material is quickly summarized in section~3.  The
analysis of anomalies for closed strings is contained in section~4; some of
those results are needed in section~5 as well.  Section~6 is a commentary, in
mathematical language, on some of the issues raised in the main part of the
paper.  First, we give a more precise description of the $A$- and $B$-fields
in terms of \v Cech theory.  (A general mathematical framework which would
apply to all occurrences of $p$-form fields in quantum field theory, string
theory, and $M$-theory is still lacking, so we settle for the \v Cech
description.)  Second, we remark on some general features of Dirac operators
on manifolds with boundary where the boundary conditions are {\it local\/}.

With great pleasure we dedicate this paper to Michael Atiyah.  His
influence is evident in every section.  Not only did he (and his
collaborators) develop topological $K$-theory and topological index theory,
which are used here to compute a subtle sign whose definition is analytic,
but he was also a pioneer in the application of these ideas to anomalies and
to other problems in quantum field theory.  Thus we hope that
the mix of mathematics and theoretical physics in this paper is an appropriate
tribute to
him.

 \head
 \S{1} Role Of The Anomaly In String Theory
\endhead
 \comment
 lasteqno 1@ 14
 \endcomment

We consider the Type~II superstring theory on a spacetime~$Y$, beginning
with the case that $B=0$.
 Recall that $Y$~is an oriented spin manifold.  The $D$-brane is an
oriented submanifold~$Q\subset Y$.  ($Q$ is oriented because in Type II
superstring theory, $D$-brane worldvolumes, being sources of Ramond-Ramond
flux, must be oriented.)  The string worldsheet $\Sigma$ is endowed with spin
structures $\alpha$ and $\beta$ for the left and right-movers.  We consider
maps of~$\Sigma $ into~$Y$ which send the boundary~$\partial \Sigma $ to~$Q$.
Our goal is to assess the well-definedness of the worldsheet path integral.
The relevant factors are
  $$ \pf(D)\cdot\exp \left(i\oint_{\partial\Sigma}A\right)  \tag{1.1} $$
  (where ${\pf}(D)$ is the Pfaffian of the world-sheet Dirac
  operator $D$ and $A$ is the ``$U(1)$ gauge field'' on $Q$).
Type II global anomaly cancellation for {\it closed} surfaces (applied to the
double of $\Sigma$, which is closed) shows that the result does not depend on
the spin structures (see Theorem 4.6).  So we may as well assume that
$\beta=\alpha$.  Once this is done, the Dirac operator becomes real.  So
$\pf(D)$ is real.  Its absolute value is well-defined, for example by zeta
function or Pauli-Villars regularization.  However, there is in general
no natural way
to define the sign of $\pf(D)$ as a number.
 We pick a particular sign, and proceed to
see if we will run into a contradiction.  For this, we consider a
one-parameter family of $\Sigma$'s, parametrized by a circle $C$.  Thus,
altogether we have a map $\phi:\Sigma\times C\to Y$, with $\phi(\partial
\Sigma\times C)\subset Q$.  The question is now whether, when one goes around
the loop $C$, $\pf(D)$ comes back to itself or changes sign.  Our main result
(\theprotag{5.6} {Theorem}) is that under going around $C$,
  $$ \pf(D)\to (-1)^\alpha \,\pf(D). \tag{1.2} $$
with
  $$ \alpha={\bigl(\partial\Sigma\times C,\phi^*(w_2(Q))\bigr)}.$$
Here $w_2(Q)$ is the second Stiefel-Whitney class of $Q$.
If $\phi:\partial\Sigma\times C\to Q$ is an embedding, we can just write
$$\alpha=\int_{\partial\Sigma\times C}w_2(Q)=(\partial \Sigma\times C,w_2(Q)).
$$
In particular, if $w_2(Q)$ is non-zero,\footnote{And can be detected
by a map from $\partial\Sigma\times C$, which is
a two-torus.  By analogy with many other problems involving global
anomalies, we expect, though we will not try to prove here, that if $w_2(Q)$
is nonzero but can only be detected by a map from a surface of higher
genus, then the same consequences will follow upon analyzing
the factorization of the string measure when $\Sigma$ breaks into
pieces, or in other words by analyzing unitarity of string scattering
amplitudes.}
$\pf(D)$ is not well-defined by  as a number.

Let $\nu $ be the normal bundle to $Q$ in $Y$.  Because $Y$ is spin
($w_1(Y)=w_2(Y)=0$) and $Q$ is oriented ($w_1(Q)=0$), the
Whitney sum formula
$$(1+w_1(Y)+w_2(Y)+\dots)
=(1+w_1(Q)+w_2(Q)+\dots)(1+w_1(\nu )+w_2(\nu )+\dots)$$
 gives $w_1(\nu )=0$
and $w_2(\nu )=w_2(Q)$.  Hence we can restate the above anomaly formulas,
for example
$$\alpha =(\partial\Sigma\times C,\phi^*(w_2(\nu ))).$$
This formulation turns out to be more natural in $K$-theory.

When $\pf(D)$ is not well-defined, the string theory is well-defined only if
the second factor in \thetag{1.1} has precisely the same ambiguity.  In other
words, $A$ must not be globally a conventional $U(1)$ gauge field, for which
the holonomy around a loop is well-defined as an element of $U(1)$.  Rather,
the holonomy around $\partial\Sigma$
$$ \exp\left(i\oint_{\partial\Sigma}A\right) $$
must be well-defined only up to multiplication by $\pm 1$, and must
change sign whenever $\pf(D)$ does.

There is another important differential-geometric object
that has the same sign ambiguity.
  Let $\omega$ be the Levi-Civita connection of the manifold
$Q$.  Its structure group is $SO(n)$, $n$ being the dimension of $Q$.
Consider the trace of the holonomy of this connection in the
spin representation $S$ of the double cover Spin$(n)$.  This is
customarily denoted
$$\Tr\,P\exp\left(\oint_{\partial\Sigma} \omega\right).\tag{1.3}$$
This holonomy is well-defined only up to sign (because
there are two ways to lift an element of $SO(n)$ to $Spin(n)$).
In going around a one-parameter family of loops, parametrized by a circle
$C$, the holonomy \thetag{1.3} is multiplied by exactly the same sign
factor that appears in \thetag{1.2}.  The upshot, then, is that
if $A$ is the right kind of geometrical object to make the worldsheet
string measure well-defined, then the product
$$\Tr\,P\exp\left(\oint_{\partial\Sigma} \omega\right)\cdot
\exp\left(i\oint_{\partial\Sigma}A\right) \tag{1.4}$$
is likewise well-defined.

What appears in \thetag{1.4} is the trace of the holonomy in
going around the loop $\partial\Sigma$, not for ordinary spinors on $Q$,
but for spinors of charge 1 with respect to $A$.
Such charged spinors are sections not of $S(Q)$, the ``spin bundle''
of $Q$, but of $S(Q)\otimes {\Cal L}$, where ${\Cal L}$ is the
``line bundle'' on which $A$ is a connection.  The meaning of the global
anomaly (for topologically trivial $B$-field) is thus that the globally defined
object is not in general $S(Q)$ or $\Cal L$, but the tensor product
$S(Q)\otimes \Cal L$.

Such a tensor product defines a so-called ``$\spinc$ structure'' of $Q$.
The global anomaly thus implies that (again, for trivial $B$-field)
$Q$ must be $\spinc$, and more specifically  a
$\spinc$ structure can be constructed from the physical data,
namely from the Levi-Civita connection $\omega$ and the field~$A$.

Since, as we have seen above, $w_2(Q)=w_2(\nu )$,
 it is equivalent to endow
$Q$ with a $\spinc$ structure or to endow its normal bundle with such
a structure.  In other words, being given a bundle $S(Q)\otimes \Cal L$
determines a bundle $S(\nu )\otimes \Cal L$, where  $S(\nu )$
are the spinors of the normal bundle.

\bigskip\noindent{\it Physical Interpretation}
\smallskip\nobreak
The phenomenon just indicated was first encountered by
hand in a very special case \cite{W3}, but it has a theoretical
interpretation that we will now recall.

Naively speaking, the conserved charges associated with wrapping
of $D$-branes in a spacetime manifold $Y$ take values in $H^*(Y;\ZZ)$,
the cohomology of $Y$.  A closer look shows, however \cite{MM,W4},
that, when the $B$-field is topologically trivial,
$D$-brane charge takes values in $K$-theory, in fact in $K(Y)$ or
$K^1(Y)$ for Type IIB or Type IIA string theory.

However, for a $D$-brane wrapped on $Q$ to define a class in $K(Y)$
(or $K^1(Y)$), its normal bundle $\nu $ must be endowed with a $\spinc$
structure. This results from a standard construction of Atiyah,
Bott, and Shapiro; see \cite{W4}, section 4.3 for an explanation
in the context of the application to string theory.
Hence, if the $K$-theory interpretation of
$D$-branes is correct, Type II  $D$-branes (at $B=0$) must be naturally
endowed with a $\spinc$ structure on the normal bundle.
Equivalently, since $w_2(Q)=w_2(\nu )$, the Type II $D$-brane world-volume
$Q$ must carry a $\spinc$ structure.  As we have just explained,
the global anomaly formula \thetag{1.2} has exactly this consequence;
because of this anomaly, a $D$-brane is endowed not with a $U(1)$ gauge
field, as it naively appears, but with a $\spinc$ structure.

\def\underarrow#1{\vbox{\ialign{##\crcr$\hfil\displaystyle
 {#1}\hfil$\crcr\noalign{\kern1pt\nointerlineskip}$\longrightarrow$\crcr}}}%
Let us now analyze the conditions under which $Q$ admits a $\spinc$
structure.   Consider
the exact sequence of abelian groups
$$ 0\to \ZZ\underarrow{2}\ZZ\underarrow{r}\ZZ_2\to 0,    \tag{1.5}$$
with the first map being multiplication by 2 and the second reduction modulo
2.  This short exact sequence leads to a long exact cohomology sequence:
  $$ \dots H^2(Q;\ZZ)\underarrow{r}H^2(Q;\ZZ_2)\underarrow{\beta}
     H^3(Q;\ZZ)\to\dots \tag{1.6} $$
The object $\beta\bigl(w_2(Q)\bigr)\in
H^3(Q;\ZZ)$ is the third Stiefel-Whitney class $W_3(Q)$.

Vanishing of $W_3(Q)$ is equivalent to $Q$ admitting $\spinc$ structure.
Indeed, exactness of \thetag{1.6} says that $w_2(Q)=r(x)$ for some $x\in
H^2(Q;\ZZ)$ if and only if $W_3(Q)=0$.  When there is no two-torsion in
$H^2(Q;\ZZ)$, there is precisely one $\spinc$ structure on $Q$ for every such
$x$.  In fact, given a $\spinc$ connection $\omega+A$, if ${\Cal L}$ is the
``line bundle'' on which $A$ is a connection, then ${\Cal M}={\Cal L}^2$ is
an honest line bundle, and $x=c_1({\Cal M})$ obeys $r(x)=w_2(Q)$.
Conversely, given $x$ with $r(x)=w_2(Q)$, we let ${\Cal M}$ be a complex line
bundle with $c_1({\Cal M})=x$, and we let $A$ be a connection on the ``line
bundle'' ${\Cal L}={\Cal M}^{1/2}$.  (When there is two-torsion in
$H^2(Q;\ZZ)$, there are different square roots of ${\Cal M}$ and hence more
than one $\spinc$ structure for given $x$.)

\bigskip\noindent{\it Inclusion Of $B$-Field}
\smallskip\nobreak
Now let us discuss what to do when $W_3(Q)\not=0$, so that the anomaly
cannot be cancelled by picking a  $\spinc$ structure on $Q$.
So far we have assumed that the $B$-field is topologically trivial,
in which case it can simply be ignored in analyzing the anomalies.
But now we must include it.  The $B$-field couples to the world-sheet $\Sigma$
in bulk, and in its presence an additional term must be added to
\thetag{1.4}, which now becomes
$$\pf(D)\,\exp \left(i\oint_{\partial\Sigma}A +i\int_\Sigma B\right).
\tag{1.7}$$
This expression has the gauge invariance
$$A\to A-\Lambda,~~ B\to B+d\Lambda, \tag{1.8}$$
where $\Lambda$ is any connection on an arbitrary complex line bundle ${\Cal
M}$.

Let us discuss, in stages, the meaning of \thetag{1.7}.  First
we consider the $B$-field in the theory of closed oriented bosonic
strings.
In this theory, for a closed surface $\Sigma$, the $B$-field gives
a phase
$$W(\Sigma;B)=\exp\left(i\int_\Sigma B\right). \tag{1.9}$$
This is a complex number of modulus one, an element of $U(1)$; we can think
of it as the holonomy of $B$ over $\Sigma$.  On the other hand, if $\Sigma$
has a boundary, then $W(\Sigma;B)$ is not gauge-invariant.  It must be
regarded as an element of a complex line ${\Cal L}_B$ associated to $\partial
\Sigma$ by $B$.  The line~${\Cal L}_B$ depends on $\partial\Sigma$, but we do
not show this in the notation.  For example, if $\partial\Sigma$ is a single
circle, then ${\Cal L}_B$ varies as $\partial\Sigma$ varies to give a complex
line bundle over the loop space of $Y$ (or over a component of this loop
space determined by the homotopy class of $\Sigma$).  We write $LY$ for the 
loop
space of $Y$.

The interpretation of $W(\Sigma;B)$ as taking values in a complex
line ${\Cal L}_B$ may seem slightly abstract, but it actually reflects
an idea that is familiar to physicists.  The two-form field $B$ in spacetime
determines (by integration over the loop) a one-form field or abelian
gauge field on loop space; this abelian gauge field is a connection
on a complex line bundle ${\Cal L}_B$ over $LY$.

Now we introduce in the bosonic string a $D$-brane with world-volume
$Q$. We require $\partial\Sigma\subset Q$.  In this situation,
 there is a completely gauge-invariant extension of
\thetag{1.9}, namely
$$W(\Sigma;B,A)=
\exp\left(i\oint_{\partial\Sigma}A+i\int_\Sigma B\right). \tag{1.10}$$
We interpret this expression to mean that $A$ gives a trivialization of
the restriction of ${\Cal L}_B$ to loops that lie in $Q$.  In other
words, $A$ trivializes the restriction of ${\Cal L}_B$
to $LQ$, the loop space of $Q$; when $\partial\Sigma\subset Q$,
$W(\Sigma;B,A)$ is a gauge-invariant version of $W(\Sigma;B)$.

Now, let us consider the superstring and the problem of defining
the product
$$ {\pf}(D)\cdot \exp\left(i\int_\Sigma B\right)\cdot \exp\left(i\oint
_{\partial \Sigma}A\right). \tag{1.11}$$
For simplicity of exposition in what follows,
we suppose that the boundary of $\Sigma$ is a single
circle; the generalization is evident.
In the mathematical theory, ${\pf}(D)$, though not well-defined
as a number, is defined as a section of a Pfaffian line bundle
${\Pf}$ over $LQ$.  As we have already discussed, the second
factor in \thetag{1.11} must likewise be interpreted as a section of a line
bundle ${\Cal L}_B$.  We also have just explained that, in the bosonic
string theory, the last factor in \thetag{1.11} should be interpreted
as trivializing the restriction of ${\Cal L}_B$ to $LQ$.
For the superstring, since there is an extra factor in \thetag{1.11},
the interpretation is different.  The last factor in \thetag{1.11} must
in this case be understood as giving a trivialization of
${\Pf}\otimes {\Cal L}_B$, restricted to $LQ$.

This means in particular that for the bosonic string, the restriction
of $B$ to $Q$ must be topologically trivial.  For the superstring,
the restriction of $B$ to $Q$ is in general topologically nontrivial, but
its topological type is uniquely determined, by the fact that ${\Pf}\otimes
{\Cal L}_B$
(restricted to loops in $Q$) must be topologically trivial.

\bigskip\noindent{\it Topological Classification Of $B$-Fields}

To make this somewhat more explicit, we recall the topological classification
of $B$-fields.  Topologically, $B$-fields on $Y$ are classified by a
characteristic class $\zeta\in H^3(Y;\ZZ)$.  At the level of differential
forms, $\zeta$ is represented by $H/2\pi$, where $H=dB$ is the curvature of
$B$.  By integrating $\zeta$ over a loop, we get a two-dimensional
characteristic class on $LY$, which in fact equals $c_1({\Cal L}_B)$.  {\it
Flat\/} $B$-fields, that is, $B$-fields whose field strength~$H$ vanishes,
are classified by the holonomy around closed surfaces, which gives a
cohomology class in~$H^2(Y;\RR/\ZZ)$.  A flat $B$-field has a torsion
characteristic class~$\zeta $ computed by the ``Bockstein'' map $\beta
\:H^2(Y;\RR/\ZZ)\to H^3(Y;\ZZ)$.  (A similar~$\beta $ appears
in~\thetag{1.6}.)  Notice by comparison the analogous classification of
abelian gauge fields, where the degree is shifted down by one.

$\Pf\otimes {\Cal L}_B$  is trivial if
\footnote{In fact, \thetag{1.12}~is only a sufficient condition,
but is the only cohomological condition that implies triviality
of $\Pf\otimes {\Cal L}_B$.
As in footnote~2 at the beginning of section~1, we expect
that factorization and unitarity will lead to
the full requirement of~\thetag{1.12}.} 
  $$ \zeta\vert_Q=W_3(Q),  \tag{1.12}$$
  and we claim that this is the right topological condition on the
  $B$-field.
The integral of the left hand side over~$\bS$ is the first Chern class
of~$\scrL_B$.  The holonomy of~$\Pf$ was stated in~\thetag{1.2}, and this may
be reinterpreted to say that the class of $\Pf$ 
as a {\it
flat\/} line bundle
in~$H^1(LQ;\RR/\ZZ)$  is the integral of~$w_2(Q)$ over~$\bS$.  Using the remark
following~\thetag{1.6} and the fact that the Bockstein commutes with
integration, we see that the first Chern class of~$\Pf$ is the integral
of~$W_3(Q)$ over~$\bS$.

In particular, $B$ restricted to $Q$ is topologically trivial if and
only if $W_3(Q)=0$, or in other words (given what we have seen above)
if and only if $Q$ is $\spinc$.
\comment
  When $Q$ is $\spinc$, ${\Cal L}_B$ is
topologically trivial.  Its trivializations are $\spinc$ structures on $Q$.
{\it is it possible to give a simple explanation of this assertion?}
Hence, we recover our earlier assertions that when $B$ is topologically
trivial, $Q$ must be $\spinc$ and $A$ is a $\spinc$ structure on $Q$.
\endcomment

To be explicit, we will give an example of a $B$-field
on $Q$ (or on any spacetime $Y$ over which $w_2(Q)$ extends)
with $\zeta=W_3(Q)$.
Just as an abelian gauge field is completely
determined up to isomorphism by its holonomy around closed loops, a $B$-field
is completely determined up to isomorphism by its holonomy over closed
surfaces.  Since $W_3(Q)=\beta(w_2(Q))$
is a torsion class,  there exists a flat
$B$-field, whose holonomy over a closed surface $\Sigma$ depends only
on the homology class of $\Sigma$, with $\zeta=W_3(Q)$.  Such
a flat $B$-field is indeed described by the elegant formula
  $$ W(\Sigma;B)= (-1)^{(\Sigma,w_2(Q))} . \tag{1.13}$$
In other words, the isomorphism class of this flat $B$-field is the image
of~$w_2(Q)$ under $H^2(Q;\ZZ_2)\to H^2(Q;\RR/\ZZ)$.  Its characteristic
class~$\zeta $ is then computed by the Bockstein map to be~$W_3(Q)$.

\bigskip\noindent{\it Examples}
\smallskip\nobreak
We will give a few concrete examples to which the discussion applies.

Every oriented manifold of dimension $\leq 3$ is spin.  A simple
example of a four-manifold that is not spin is
 $Q=\CC P^2$.  Then $H^2(Q;\ZZ)=\ZZ$, and $H^2(Q;\ZZ_2)=\ZZ_2$.  The second
Stiefel-Whitney class~$w_2(Q)$ is the nonzero element of $H^2(Q;\ZZ_2)$.
Since $H^3(Q;\ZZ)=0$, we have $W_3(Q)=0$ and $Q$ is $\spinc$.  In fact, the
map $r:H^2(Q;\ZZ)\to H^2(Q;\ZZ_2)$ is just reduction modulo 2; the elements
$x\in H^2(Q;\ZZ)$ with $r(x)=w_2(Q)$ correspond to the odd integers in
$H^2(Q;\ZZ)\cong \ZZ$.

If a $D$-brane has world-volume $Q=\CC P^2$, then (assuming that the
$B$-field vanishes)  the global anomaly means
that the ``$U(1)$ gauge field'' $A$ does not obey standard Dirac quantization.
Rather, $A$ is a connection on a ``line bundle'' whose {\it square}  has an odd
 first Chern class $x$ (congruent to $w_2(Q)$ mod 2).
If $L\subset Q$ is a generator of  $H_2(Q;\ZZ)$, then we have
$(L,w_2(Q))\not= 0$ and hence
$$\int_{L}{F_A\over 2\pi}={x\over 2},$$
with $x$ an odd integer.
In particular, $x$ cannot be zero.

Every oriented manifold of dimension $\leq 4$ is $\spinc$.  To give
a simple example of a five-manifold that is not $\spinc$, let
$Q'=Q\tilde \times \SS^1$ be a $\CC P^2$ bundle over $\SS^1$ in which,
as one goes around the $\SS^1$, the fiber $\CC P^2$ undergoes complex
conjugation.  Complex conjugation acts on $H^2(Q;\ZZ)$ by multiplication
by $-1$.  $Q'$ is not a $\spinc$ manifold, for the following reason.
An $\tilde x\in H^2(Q';\ZZ)$ obeying $r(\tilde x)=w_2(Q')$ would have to reduce
on each fiber of the projection $Q'\to \SS^1$ to an odd element
$x\in H^2(Q;\ZZ)\cong \ZZ$.  Because of the monodromy, $x$ would
have to change sign in going around the $\SS^1$, which is impossible.
So $Q'$ is not $\spinc$.

To illustrate the ideas of the present paper requires considering
topologically non-trivial $D$-brane world-volumes such as $Q$ or $Q'$.
But spacetime itself can be very simple, for example $Y=\RR^{10}$.  Since
$Y$ is contractible, the $B$-field is automatically topologically
trivial and can be ignored. We will give examples of $D$-brane states
in $\RR^{10}$ which are or are not allowed.

On general grounds, any manifold of dimension $\leq 5$ can be embedded
in $\RR^{10}$.  For completeness, we will describe a simple
embedding of $Q$ and $Q'$.
 Identify $Q=\CC P^2$ with the space of vectors $\psi$
of unit norm in $\CC^3$, modulo phase rotations.  Let $\lambda_a$
be a suitably normalized basis of the Lie algebra of $SU(3)$, and
$$\phi_a=\langle\psi|\lambda_a|\psi\rangle.$$
Then $\sum_{a=1}^8\phi_a^2=1$, so using the $\phi_a$ as coordinates,
we get an embedding of $\CC P^2$ in $\SS^7$, which obviously embeds
in $\RR^{10}$.  To embed $Q'$ in $\RR^{10}$, first note that
$\SS^8\times \SS^1$ can be embedded in $\RR^{10}$ (for instance
as the set of points a distance $\epsilon$ from a circle $\SS^1\subset
\RR^{10}$, for suitably small $\epsilon$).  Hence it suffices to
embed $Q'$ in $\SS^8\times \SS^1$.  For this, we first add one more
coordinate $y$ to the $\phi_a$, to get a copy of $\SS^8$ defined by
$\sum_a\phi_a^2+y^2=1$, and embed $Q$ in $\SS^8$ as the subset
with $y=0$ and $\phi_a$ as before.  Now, complex conjugation
of $\CC P^2$ acts on the set of eight $\phi_a$ with determinant $-1$
(three eigenvalues $+1$ and five $-1$), so if we take it to act on $y$
as multiplication by $-1$, we get an element $T\in SO(9)$.  $SO(9)$  is
a connected group of symmetries of $\SS^8$.  Let
$R(\theta)$, $0\leq \theta\leq 2\pi$ be a path in $SO(9)$ with
$R(0)=1$ and $R(2 \pi)=T$.  Finally, embed $Q'$ in $\SS^8\times
\SS^1$ by mapping $(P,\theta)$ (with $P\in \CC P^2\subset \SS^8$ and
$\theta\in \SS^1$) to $(R(\theta)P,\theta)\in \SS^8\times \SS^1$.

Now, given a $D$-brane in $\RR^{10}$ with world-volume $Q$ or $Q'$,
we will explain how to construct a family of string worldsheets
that detects the global anomaly.  We take $\Sigma$ to be a disc; its
boundary is a circle $\partial \Sigma$.
Letting $C$ be another circle,
the boundary of $\Sigma\times C$ is $W= \partial\Sigma\times C$, a two-torus.
For a generator of $H_2(Q;\ZZ)$ or $H_2(Q';\ZZ)$ we can take
a two-sphere $L$.  We pick a degree one map $\phi_0:W\to L$; then, since
 $\RR^{10}$ is contractible, one can extend $\phi_0$ to
a map $\phi:\Sigma\times C\to \RR^{10}$.
This gives a relatively simple example of a
 family of worldsheets for which there is a global anomaly.
Hence, the $D$-brane $Q'\subset \RR^{10}$ is not allowed,
and the $D$-brane $Q\subset \RR^{10}$ is allowed but must
support a ``$U(1)$ gauge field'' with half-integral flux.

\newpage\noindent
{\it Conservation Laws And The Anomaly}
\nobreak\smallskip\nobreak
By further discussion of $D$-branes in $\RR^{10}$, we can show
the relation of the anomaly to $D$-brane charges.

We work in Lorentz signature, and split $\RR^{10}$ as $\RR\times \RR^9$,
where $\RR$ parametrizes ``time'' and $\RR^9$ is ``space.'' We consider
a $D$-brane whose world-volume near time zero is approximately
$Q=\RR\times Q_0$,
with $Q_0\subset \RR^9$.  We will focus on Type IIA superstrings, so
$Q_0$ is of even dimension.  To detect the anomaly, the dimension is
at least four.

The world-volume of the $D$-brane will not look like $\RR\times Q_0$ for
all time.  The $D$-brane will oscillate, emit radiation, and contract.
In the far future, it will decay to a final state consisting
of outgoing stable particles.  The only known stable particles in Type  IIA
superstring theory in $\RR^{10}$ are massless particles (the graviton
and its superpartners) and $D0$-branes, together with the familiar
multi-$D0$-brane bound states.   (There are no conserved
charges for higher branes because $\RR^{10}$ is contractible.)
If there
are no additional stable particles, then our initial state will decay
to an assortment of the known ones.

If so, we can predict how many net $D0$-branes will be produced --
that is the difference between the number of $D0$-branes
and anti-$D0$-branes in the final state.   It must equal the
$D0$-brane charge of the initial state, which \cite{GHM,CY,MM} is
$$N_0=\int_{Q_0}\sqrt{\hat A(Q)}{1\over \sqrt{\hat A(\nu )}}\,\exp
\left({c_1({\Cal L})}\right),$$
where $\hat A$ is the total A-roof class, and $c_1({\Cal L})$ is the
first Chern
class of the ``complex line bundle''  ${\Cal L}$ on which the ``$U(1)$
gauge field'' $A$ of the $D$-brane is a connection.  Now, using
the fact that the tangent bundle of $\RR^{10}$ is trivial, and splits
as $TQ\oplus \nu $ (with $TQ$ the tangent bundle to $Q$), we have
$\hat A(\nu )=\hat A(Q)^{-1}$.  So we rewrite the formula for the total
$D0$-brane charge as
$$ N_0 =\int_{Q_0}\hat A(Q_0)\,\exp\left({c_1({\Cal L})}\right). \tag{1.14} $$
(We have written here $\hat A(Q_0)$ rather that $\hat A(Q)$; the
two are equal as $TQ=TQ_0\oplus \epsilon$, where $\epsilon$ is a trivial
real line bundle that incorporates the ``time'' direction.)

Now, if ${\Cal L}$ were a complex line bundle, then in general $N_0$
would not be an integer.  For example, for $Q_0=\CC P^2$ and
$\Cal L$ trivial,
we would have $N_0=\pm 1/8$
(depending on orientation).  When $N_0$ is not
integral, the initial $D$-brane state
cannot  decay to known stable particles.

We either must postulate the existence of new conservation laws
for Type IIA strings in $\RR^{10}$ or achieve integrality of $N_0$
by some other modification of the rules.  As we have seen, the global
anomaly means that $A$ is not a $U(1)$ connection, but rather determines
a $\spinc$ structure $S(Q)\otimes {\Cal L}$ on $Q$ or equivalently a $\spinc$
structure $S(Q_0)\otimes {\Cal L}$ on $Q_0$.  With this interpretation of
$A$ and ${\Cal L}$, the Atiyah-Singer index theorem states
that the right hand side of \thetag{1.14} is the index of the Dirac
operator (for spinors on $Q_0$ valued in $S(Q_0)\otimes {\Cal L}$),
so that $N_0$ is always integral.

Thus, the anomaly
enables us to avoid having to postulate new conservation
laws for $D$-branes in $\RR^{10}$.

\head
\S{2} Qualitative Explanation Of Anomaly
\endhead
 \comment
 lasteqno 2@000
 \endcomment

To give a qualitative explanation of the anomaly,
we consider a $Dp$-brane with oriented world-volume $Q$ in a spacetime
$Y$.  If the given $D$-brane is the only one in spacetime, then a
consideration of the string spectrum will not lead in an obvious way to a
result involving $w_2(Q)$.  In fact, if there is only one $D$-brane, the only
open strings are the $p$-$p$ open strings with both ends on $Q$.  The ground
state of the $p$-$p$ open strings in the Ramond sector consists of sections
of $S(Q)\otimes S(\nu )$ where $S(Q)$ is the bundle of spinors on $Q$, and
$S(\nu )$ is the bundle of spinors on the normal bundle $\nu $ to $Q$ in $Y$.
This tensor product exists whether $Q$ is spin or not (given only that $Y$ is
spin), so merely by quantizing the $p$-$p$ open strings, we get no condition
involving $w_2(Q)$.  To obtain such a condition, we will study global
worldsheet anomalies, as explained in the introduction.

Suppose, however, that an additional pair of $D$-branes, consisting
of a space-filling $9$-$\bar 9$ pair, is present.\footnote{For Type IIB,
these can be ordinary supersymmetric branes.  For Type IIA,
we could in this argument use instead the nonsupersymmetric 9-brane
considered in \cite{H}.}
  We suppose that the
gauge fields on the $9$-brane and $\bar 9$-brane are trivial.
In the presence of the additional branes, there are additional
open strings such as the $9$-$p$ open strings.  The ground state
Ramond sector $9$-$p$ strings are sections of $S(Q)\otimes {\Cal L}$,
where again $S(Q)$ are the spinors on $Q$ and ${\Cal L}$ is the
``line bundle'' on which the ``$U(1)$ gauge field'' $A$ on $Q$ is
a connection.

So the tensor product $S(Q)\otimes {\Cal L}$ must exist, and we learn
what was promised in the introduction: $Q$ must be $\spinc$,
and the ``gauge field'' on $Q$ really defines a $\spinc$ structure.

In our actual problem, such additional branes are absent.
We do not want to assume continuous creation and annihilation
of brane-antibrane pairs along the lines of \cite{S}, for this
involves somewhat speculative physics.  If one is willing to make
such assumptions, the relation of
branes to $K$-theory and the $\spinc$ character of $Q$ can indeed be
deduced, as in \cite{W4}.  Our intent here is to show that the
requirement for $Q$ to carry a $\spinc$ structure can be deduced
with only conservative assumptions about physics, by computing
the global worldsheet anomaly.

Nevertheless, the fact that in the presence of a 9-brane, the quantization
of the $9$-$p$ open strings would make our result obvious is a starting
point for a precise mathematical computation of the global anomaly.
In essence, whether or not 9-branes can be continuously created in the
physics, we can create them in the math, at least for the purposes
of computing a global anomaly.  This may be done as follows.

Consider any family of open string worldsheets $\Sigma$ with specified
boundary conditions on the boundary components of $\Sigma$.
The change in the global anomaly under a change in the boundary
conditions is local, that is it only depends on the properties
near the boundary of $\Sigma$.  Such locality is perhaps more
familiar for perturbative anomalies, which are expressed as integrals
of differential forms.  Global anomalies, however,
also obey a suitable form of locality.  They are computed topologically
as ``integrals'' in $K$-theory, which obey the following excision
property: If two $K$-theory classes agree outside an open set $U$, then the
difference of their ``integrals'' can be computed on $U$.
Geometrically, global anomalies are adiabatic limits of $\eta$-invariants,
which also obey a factorization relation, though a more delicate
one \cite{DF}.  These relations give the locality we want for the global
anomaly.

Using this locality, we can reduce to  a convenient set of boundary
conditions.
In fact, if one places $9$-brane boundary conditions on all components of
$\partial \Sigma$,
there is no global anomaly.  This statement is proved in Proposition
5.10, again using factorization.
The statement is closely related to Type II global anomaly cancellation
for closed surfaces without $D$-branes, because with 9-brane boundary
conditions on all boundaries, the Dirac equation on the surface-with-boundary
$\Sigma$ is equivalent to a chiral Dirac equation (acting on spinors
of one chirality only) on the double of $\Sigma$.

Hence, the anomaly for the family $\Sigma\times C$ of open string world-sheets,
with a map $\phi:\Sigma\times C\to Y$, depends only on the restriction
of $\phi$ to $\partial\Sigma\times C$ and the boundary conditions on
$\partial\Sigma\times C$.  Once this is known, the anomaly can be evaluated
by the following sleight of hand.
We let $\Theta$ be an annulus.  We select one distinguished
component $\partial\Theta'$ of the boundary of $\Theta$, and we select
an isomorphism of $\partial\Theta'$ to $\partial \Sigma$ (assuming $\partial
\Sigma $~connected, in which case
both of them
are circles; the generalization to arbitrary~$\partial \Sigma $ is clear).
Then we pick a map $\tilde\phi:\Theta\times C\to Y$
which coincides with $\phi$ in a neighborhood of
 $\partial\Theta'\times C$.

Now, we will compare the global anomalies for two different sets
of Dirac operators that differ only by changes of boundary conditions:

(1)  In the first case, we consider a Dirac operator on $\Sigma$ with
$p$-brane boundary conditions, plus a Dirac operator on $\Theta$
with $9$-brane boundary conditions at each boundary.

(2) In the second case, we consider a Dirac operator on $\Sigma$
with $9$-brane boundary conditions, and a Dirac operator on $\Theta$
with $p$-brane boundary conditions on $\partial\Theta'$ and $9$-brane
boundary conditions on the other boundary component.

The total global anomaly in going around the loop $C$, summed over the
two Dirac operators that are considered in each case, is the same
in case (1) and case (2), because the union  of all the boundaries
and boundary maps are the same in the two cases.  All we have done
in going from case (1) to case (2) is to cut out neighborhoods of
boundary components of $\Sigma\times C$ and $\Theta\times C$ and exchange them.

We also observe the following:

(I)  In case (1), the global anomaly comes entirely from the
Dirac operator on $\Sigma$, since the Dirac operator on $\Theta$
has $9$-brane boundary conditions on each component.  Hence, in case (1),
the global anomaly is equal to what we want to calculate.

(II) In case (2), the global anomaly comes entirely from the Dirac
operator on $\Theta$, since the Dirac operator on $\Sigma$ has $9$-brane
boundary conditions on each component.

In case (2), because $\Theta$ is an annulus with $9$-brane boundary
conditions at one end and $p$-brane boundary conditions at the other,
it describes the propagation of $9$-$p$ strings.  As we remarked before,
the ground states of such strings  are sections
of $S(Q)\otimes {\Cal L}$  in the Ramond sector
(or of  $S(\nu )\otimes {\Cal L}$ in the Neveu-Schwarz sector).
Absence of global anomalies in their
propagation is equivalent (as in the analysis of global anomalies
in quantum mechanics in \cite{W2})
to the existence of $S(Q)\otimes {\Cal L}$ (or of $S(\nu )\otimes {\Cal L}$).

Thus, absence of the global anomaly for the Dirac operator on $\Sigma$
with $p$-brane boundary conditions is equivalent to existence
of $S(Q)\otimes {\Cal L}$, as we wished to show.  A
rigorous argument  can be found in section 5.

\head \S{3} Anomalies and Index Theory
 \endhead
 \comment
 lasteqno 3@  7
 \endcomment

The path integral over a fermionic field~$f$ is the regularized pfaffian,
or\footnote{In Minkowski space all fields are real and we can write the
fermionic kinetic term as~$\frac 12\psi \Dirac\psi $, so that the path
integral over~$\psi $ gives a real pfaffian.  Often a pfaffian may be written
as a determinant of a ``smaller'' Dirac operator.  Since under Wick rotation
the fermions are usually complexified, in Euclidean field theory the
pfaffians and determinants are usually complex.} determinant, of a Dirac
operator~$D$.  It depends on the bosonic fields~$b$ which couple to~$f$, but
rather than being a complex-valued function on the space~$B$ of these bosons
it is a section~$\pf D$ of a complex line bundle~$\Pf D\to B$.  This section
is part of the effective action we must integrate over~$B$, and to do so we
must find a global nonzero section~$\bo\:B\to\Pf D$ and integrate instead the
function~$\pf D/\bo$.  The anomaly is the obstruction to finding the
trivializing section~$\bo$.  (More precisely, we must trivialize the product
of the pfaffian line bundle and line bundles from other terms in the
effective action, such as the additional contributions to~\thetag{1.11}.)
This may be interpreted topologically, in which case the Atiyah-Singer index
theorem is used to determine the topology of the pfaffian line bundle.  More
relevant to the physics is a geometric interpretation, in which we seek a
flat section~$\bo$ of unit norm relative to the natural metric and connection
on~$\Pf D$.  (This distinction is important in our problem---compare
\theprotag{5.5} {Theorem} and \theprotag{5.6} {Theorem}.)  For that we use
differential geometric index theorems involving curvature forms and $\eta
$-invariants.  If the anomaly vanishes then~$\bo$, and so the effective
action, is determined up to a phase on each connected component of~$B$.
See~\cite{F1}, ~\cite{F2} for more details.  In this section we briefly
summarize results from index theory---both topological and geometric---in the
form we need.

Suppose $\pi \:X\to Z$ is a fiber bundle whose fibers\footnote{Here
$X/Z$~denotes the fiber of $X\to Z$.  As usual, a spin structure on a
manifold means a spin structure on its tangent bundle, here the tangent
bundle~$T(X/Z)$ along the fibers.}~$X/Z$ are closed manifolds endowed with a
spin structure.  Recall that $KO(X)$~is the group of virtual real vector
bundles on~$X$ up to equivalence, and $KO^{-n}(X)\subset KO(S^n\times X)$ is
the subspace of isomorphism classes of virtual bundles trivial on~$S^n\vee
X$.  The spin structure on the fibers determines a pushforward map
  $$ \pi _!^{X/Z}\:KO(X)\longrightarrow KO^{-n}(Z),  $$
where $n=\dim X/Z$.  Given a real vector bundle $E\to X$, the family of Dirac
operators on~$X/Z$ coupled to~$E$ has an index in~$KO^{-n}(Z)$.  The
Atiyah-Singer index theorem~\cite{AS2} asserts that this analytic index
equals $\pi _!^{X/Z}\bigl([E] \bigr)$, where $[E]\in KO(X)$ is the
isomorphism class of~$E$.  For Dirac operators coupled to complex bundles we
have a similar picture with complex $K$-theory replacing real $KO$-theory.
In general there is no cohomological formula for the integral or mod~2
characteristic classes of the index; over the rationals we can express the
Chern character of the index (in $K$-theory) in terms of the Chern character
of~$E$:
  $$ \ch \pi _!^{X/Z}\bigl([E] \bigr) = \pi _*^{X/Z}\bigl(\Ahat(X/Z)\ch(E)
     \bigr),  $$
where $\pi _*$~is the pushforward map in rational cohomology.  Until further
notice we restrict to complex bundles and $K$-theory.

The simplest invariant of an element of~$K(Z)$ after the rank, which is a
continuous function $Z\to \ZZ$, is the {\it determinant line bundle\/}, which
is a smooth complex line bundle over~$Z$.  In this topological context it is
only defined up to equivalence.

A {\it geometric\/} family of Dirac operators parametrized by~$Z$ is
specified by a fiber bundle~$\pi\:X\to Z$, a spin structure on~$X/Z$, a
Riemannian metric on~$X/Z$, and a distribution of horizontal planes on~$X$
(transverse to the fibers~$X/Z$).  If we couple to a vector bundle $E\to X$,
we require that $E$~have a metric and compatible connection.  Then if the
fibers~$X/Z$ are closed, the determinant line bundle $\Det D^{X/Z}(E)$ is
well-defined as a smooth line bundle, and it carries a canonical metric and
connection~\cite{BF}.  If the fibers~$X/Z$ are odd dimensional, so that the
(complex) Dirac operator is self-adjoint, then there is a geometric invariant
$\xi _{X/Z}(E)\:Z\to \RR/\ZZ$ defined by Atiyah-Patodi-Singer.  (It is half
the sum of the $\eta $-invariant and the dimension of the kernel.)
Multiplying by~$2\pi \sqrt{-1}$ and exponentiating we obtain $\tau
_{X/Z}(E)\:Z\to\TT$, where $\TT\subset \CC$ is the unit circle.

The curvature of the determinant line bundle is the 2-form
  $$ \Omega ^{\Det D^{X/Z}(E)} = \Bigl[2\pi \sqrt{-1}\,\int_{X/Z}
     \Ahat(\curv{X/Z})
      \ch(\curv{E})\Bigr]_{(2)}\in \Omega ^2(Z), \tag{3.1} $$
where $\curv{X/Z}, \curv{E}$  are the indicated curvature forms.  As for the
holonomy, consider a loop $\pi \:X\to S^1$ of manifolds in this geometric
setup.  Endow~$\cir$ with a metric and the {\it bounding\/} spin structure;
then we induce a metric and spin structure on~$X$.  The holonomy of the
determinant line bundle around this loop is
  $$ \hol\Det D^{X/S^1}(E) = \alim \tau \inv _X(E), \tag{3.2} $$
where the adiabatic limit~$\alim$ is the limit as the metric on~$\cir$ blows
up ($g_{\cir}\to g_{\cir}/\epsilon ^2$ and $\epsilon \to0$).  If the
determinant line bundle is flat, then no adiabatic limit is required.
Equation~\thetag{3.2} is the global anomaly formula~\cite{W1}; cf.~\cite{BF}.

If $X$~is a spin manifold of odd dimension~$n$, and $F\to X$ a {\it flat\/}
unitary bundle of rank~$r$, then for any complex hermitian bundle $E\to X$
with connection the ratio $\tau _X\bigl(E\otimes (F-r)\bigr)/\tau _X(E)$ is a
topological invariant independent of the geometrical quantities.  The flat
index theorem of Atiyah-Patodi-Singer~\cite{APS2} gives a $K$-theory formula
for this ratio.  Namely, the bundle~$F$ determines a class $[F-r]\in K\inv
(Z;\RR/\ZZ)$ and the difference of $\xi $-invariants is $\pi
_!^X\bigl([E]\cdot [F-r] \bigr)$, where now
  $$ \pi _!^X\:K\inv (Z;\RR/\ZZ)\to K^{-n-1}(\pt;\RR/\ZZ)\cong
     \RR/\ZZ. \tag{3.3} $$

If $X$~is an odd dimensional spin manifold with boundary, one can still
define~$\tau _X(E)$, but now it is an element of the inverse determinant line
of the boundary, viewed as a $\zt$-graded one dimensional vector
space~\cite{DF}.  Here we use the global boundary conditions of
Atiyah-Patodi-Singer~\cite{APS1}.  The invariant~$\tau _X(E)$ satisfies a
gluing law.  We need the formula for~$\tau _{-X}$, where $-X$~is the
manifold~$X$ with the opposite orientation.\footnote{Unfortunately, the sign
in~\thetag{3.4} is missing from~\cite{DF}; it will be corrected in a
forthcoming {\it erratum\/}.}  Let $k$~be the number of components of~$\bX$
on which the boundary Dirac operator has odd index.  Then
  $$ \tau _{-X}(E) = (-1)^{k\choose 2}\tau _X(E). \tag{3.4} $$
Let $\Xd$~be the (spin) double of~$X$, obtained by gluing~$X$ and~$-X$
along~$\bX$.  Then the gluing theorem and~\thetag{3.4} imply
  $$ \tau _{\Xd} = (-1)^{k\choose2}. \tag{3.5} $$

Turning now to real bundles and $KO$-theory, there is a square root which one
can canonically extract in certain dimensions (see~\cite{F1,\S3}).  Namely,
if a geometric family $\pi \:X\to Z$ has $\dim X/Z\equiv 2\pmod8$, and if
$E\to X$ is a real vector bundle, then the determinant line bundle has a
natural square root, the {\it pfaffian line bundle\/}~$\Pf D^{X/Z}(E)$.
Also, if $X$~is a closed spin manifold with $\dim X\equiv 3 \pmod8$, and if
$E\to X$ is a real vector bundle, then $\tau _X(E)\in \TT$ has a natural
square root $\tau _X^{1/2}(E)\in \TT$.  The curvature of the pfaffian line
bundle is given by one-half times~\thetag{3.1}, and the holonomy
by~\thetag{3.2} with the $\tau ^{-1/2}$-invariant replacing the $\tau\inv
$-invariant.  Formulas~\thetag{3.4} and~\thetag{3.5} hold for the $\tau
^{-1/2}$-invariant, but now $k$~denotes the number of components of~$\bX$
with nonzero mod~2 index.  (Note that the $\tau ^{1/2}$-invariant of a
manifold with boundary lives in the inverse pfaffian line of the boundary.
That pfaffian line is $\zt$-graded by the mod~2 index.)  There is a version
of the flat index theorem~\thetag{3.3} which applies to this square root; it
uses~$KO$ in place of~$K$.

There is a special low dimensional situation in which the topological
isomorphism class of the pfaffian line bundle (over the integers) is computed
by a cohomological formula.  Suppose $\pi \:X\to Z$ is a family of closed
spin 2-manifolds, and $E\to X$ a virtual real vector bundle of rank~0 which
is endowed with a spin structure.  Then~\cite{F1,\S5}
  $$ c_1\Pf D^{X/Z}(E) = \pi _*^{X/Z}\lambda (E), \tag{3.6} $$
where $\lambda $~is the degree four characteristic class of spin bundles with
$2\lambda =p_1$.  As a corollary, even if $E$~is not spin we have a
cohomological formula for the determinant line bundle, which is the square of
the pfaffian bundle:
  $$ c_1\Det D^{X/Z}(E) = \pi _*^{X/Z}p_1 (E). \tag{3.7} $$
In~\thetag{5.22} we compute an analogous formula for the pfaffian line bundle
of a family of Dirac operators on the circle.

 \head
 \S{4} Closed Superstrings
 \endhead
 \comment
 lasteqno 4@ 17
 \endcomment

\noindent{\it Statement of Results}
\smallskip

In this section we study a closed superstring propagating in a curved
background Riemannian manifold~$Y$.  Assume that $Y$~is an oriented manifold.
Let $\Sigma $~be an oriented closed surface.  We need not assume that either
$Y$~or $\Sigma $~is connected.  The space of bosons is the product
  $$ B = \Met(\Sigma )\times \Map(\Sigma ,Y)  $$
of Riemannian metrics~$g$ on~$\Sigma $ and maps $\phi \:\Sigma \to Y$.  A
metric~$g$ on~$\Sigma $ induces a complex structure, since $\Sigma $~is
assumed oriented, and so a $\dbar$~operator $\dbar_g\:\Omega ^{0,0}\to \Omega
_g^{0,1}$.  Let $K_g$~denote the canonical bundle, which is the cotangent
bundle~$T^*\Sigma $ viewed as a complex line bundle.  A spin
structure~$\alpha $ on~$\Sigma $, which we can describe on the oriented
surface independent of any metric, gives rise to a complex line
bundle~$H_{\alpha,g }$ which satisfies $H_{\alpha,g }^{\otimes 2}\cong K_g$.
It is natural to denote~$H_{\alpha,g }$ as~$K^{1/2}_{\alpha,g }$.  Spinor
fields are sections of~$K^{1/2}_{\alpha ,g}\oplus \overline{K^{1/2}_{\alpha
,g}}$.  The chiral Dirac operators on~$\Sigma $ may be expressed in terms of
the $\dbar$~operator and its conjugate:
  $$ \alignedat2
      D^+_{\alpha ,g} &= \hphantom{-}\dbar_g &&\: \Omega
     ^{0,0}(K^{1/2}_{\alpha ,g})\longrightarrow \Omega
     _g^{0,1}(K^{1/2}_{\alpha ,g}) \\
      D^-_{\alpha ,g} &= -\partial_g &&\: \Omega
     ^{0,0}(\overline{K^{1/2}_{\alpha ,g}})\longrightarrow \Omega _g
     ^{0,1}(\overline{K^{1/2}_{\alpha ,g}}) \endaligned \tag{4.1} $$
Note that $D^- = -\overline{D^+}$.  Also, $D^+=-(D^+)^*$~is a complex
skew-adjoint operator.  For this we identify $\Omega ^{0,1}(K^{1/2})$ as the
dual space to~$\Omega ^{0,0}(K^{1/2})$: the duality pairing is pointwise
multiplication followed by integration.  The determinant of a skew-adjoint
operator has a canonical square root, the pfaffian, as explained in~\S{3}.
For any vector bundle with connection~$E\to\Sigma $ we can form the coupled
Dirac operators~$D^{\pm}(E)$.  The coupled operator is skew-adjoint if $E$~is
real.

For a line bundle~$\scrL$ we denote the dual by~$\scrL\inv $ and the
$n^{\text{th}}$~tensor power~$\scrL^{\otimes n}$ by~$\scrL^n$.  Thus
$K_g\inv $ is the holomorphic tangent bundle of~$\Sigma $ and
$K^{-3/2}_{\alpha ,g}$ is the $3^{\text{rd}}$~power of the dual
to~$K^{1/2}_{\alpha ,g}$.

Fix a spin structure~$\alpha $.  Then for a pair~$(g,\phi )\in B$ define the
complex line
  $$ L^{+}_\alpha (g,\phi ) = \Pf D^+_{\alpha ,g}(\phi ^*TY) \otimes \bigl[
     \Det D^+_{\alpha ,g}(K_g\inv )\bigr]^{\otimes(-1)} \otimes \Det
     \dbar_g(K_g\inv ) \tag{4.2} $$
and its complex conjugate
  $$ L^{-}_\alpha (g,\phi ) = \Pf D^-_{\alpha ,g}(\phi ^*TY) \otimes \bigl[
     \Det D^-_{\alpha ,g}(\Kbar_g\inv )\bigr]^{\otimes (-1)} \otimes
     \Det \partial_g(\Kbar_g\inv ). \tag{4.3} $$
As $(g,\phi )$~vary these define smooth complex line bundles~$L^{\pm}_\alpha
\to B$ with metric and connection.  We remark that the last operator
in~\thetag{4.2} may be rewritten in terms of spinors:
  $$ \dbar_g(K_g\inv ) = D^+_{\alpha ,g}(K^{-3/2}_{\alpha ,g})  $$
for any spin structure~$\alpha $.

In superstring theory, the
spin structures on right- and left-movers are chosen
independently and are summed over with appropriate weighting.
 The path integral over the right-moving fermions is a
section of~$L^+_\alpha$,
 and the path integral over the left-moving fermions is a
section of~$L^-_\beta$, for two independently chosen spin structures
$\alpha$ and $\beta$.    The first
factor in the definition of $L^\pm_\alpha$
 corresponds to the physical spinor field, the last factor to the
ghosts from gauge fixing the diffeomorphism group, and the middle factor to
the ghosts from gauge fixing the remainder of the superdiffeomorphism group.
The path integral in the effective theory is carried out over the quotient of
the space of bosons by the subgroup of superdiffeomorphisms of~$\Sigma $
which preserve the chosen spin structure.  Thus to detect possible anomalies
we consider smooth fiber bundles $\pi\:X\to Z$ with typical fiber~$\Sigma $
together with a map to~$Y$:
  $$ \CD
     X @>\phi >> Y \\
     @VV\pi V\\
     Z \endCD
      \tag{4.4} $$
Furthermore, we assume that the relative tangent bundle carries a Riemannian
metric and spin structure, and that there is a distribution of horizontal
planes on~$X$.  This is enough to define the line bundles~\thetag{4.2}
and~\thetag{4.3} with metric and connection.  In the application to physics
$Z$~maps into the quotient of~$B$ by the subgroup of diffeomorphisms fixing
the given spin structure.  Such families of surfaces are ``probes'' which
determine the structure of the line bundles~$L^{\pm}_\alpha $.  It suffices
to take $Z$~finite dimensional and compact.  The path integral over both
right- and left-moving fermions is a section of~$L^+_\alpha \otimes L^-_\beta
$, and it is the triviality of this line bundle over arbitrary
families~\thetag{4.4} which we investigate.

We now state the basic results for closed oriented surfaces; the proofs
follow below.  In these theorems {\it isomorphism\/} means an invertible
linear map which preserves the metric and connection.  A {\it
trivializable\/} bundle is isomorphic to the trivial bundle with product
metric and connection.

        \proclaim{\protag{4.5} {Theorem}}
 Suppose $Y$~is spin and $\dim Y=10$.  Then for any two spin
structures~$\alpha ,\beta $ there exists an isomorphism
  $$ L^+_\alpha \cong L^+_\beta .  $$
        \endproclaim

 \flushpar
 Since $L^-_\alpha $~is the complex conjugate of~$L^+_\alpha $, it follows
that $L^-_\alpha $ is also independent of the spin structure~$\alpha $.

        \proclaim{\protag{4.6} {Corollary}}
 Suppose $Y$~is spin and $\dim Y=10$.  Then $L^+_\alpha \otimes L^-_\beta $
is trivializable for any spin structures~$\alpha ,\beta $.
        \endproclaim

\flushpar
 For $\alpha =\beta $ the hermitian metric provides the desired
trivialization.  The result for~$\alpha \not= \beta $ follows immediately
from \theprotag{4.5} {Theorem}.  The triviality of $L_\alpha^+ \otimes
L_\beta ^-$ is the vanishing of the anomaly.

The next result is relevant to the conformal anomaly, but does not enter into
the considerations of~\S{5}.  We include it for completeness.

        \proclaim{\protag{4.7} {Theorem}}
 Suppose $Y=\RR^{10}$.  Then $L^+_\alpha \otimes (\Det \dbar_g)^{\otimes
(-5)}$~is trivializable.
        \endproclaim

\flushpar
 The factor of $(\Det\dbar_g)^{\otimes (-5)}$ comes from integrating the
boson~$\phi $.  (See~\cite{F2,\S2} for a discussion of the conformal anomaly
in the context of the bosonic string.)

\bigskip\noindent{\it Proofs}
\smallskip\nobreak
 The proof\footnote{For an alternative proof, see~\cite{W2}.} of
\theprotag{4.5} {Theorem} is modelled on~\cite{FM}, which treats a similar
proposition for complex vector bundles.  The real case treated here is
simpler.  First, the curvature formula~\thetag{3.1} does not depend on the
spin structure, so the ratio~$L^+_\alpha /L^+_\beta $ is flat.  Thus we must
show that the holonomy of this ratio vanishes.  Now the holonomy of a
pfaffian line bundle is given by the adiabatic limit of the exponentiated
$-\xi /2$-invariant of the Dirac operator on a 3-manifold which fibers over
the circle (see~\thetag{3.2}).  \theprotag{4.5} {Theorem} follows from a more
general statement.\footnote{It seems that there is no cohomological formula
for this ratio of holonomies (i.e., the left hand side of~\thetag{4.9}) if
we relax the hypotheses $w_1(V) = w_2(V)=0$, as we learned in a conversation
with John Morgan.  Our proof of \theprotag{4.8} {Lemma} is based on the
$KO$-theory formula~\thetag{4.11}.}

        \proclaim{\protag{4.8} {Lemma}}
  Suppose $P$~is {\it any\/} compact oriented 3-manifold with two spin
structures~$\ah$ and~$\bh$, and $V\to P$ is a real vector bundle with rank
divisible by~8 and with $w_1(V)=w_2(V)=0$.  Then
  $$ \frac{\xi _{\ah\phantom{\bh}}}2 - \frac{\xi _{\bh}}2 \equiv
     0\pmod1. \tag{4.9} $$
        \endproclaim

\flushpar
 To prove \theprotag{4.5} {Theorem} from this lemma, note first that the last
factor in~\thetag{4.2} does not depend on the spin structure.  Next, use the
fact that $D^+ = \overline{D^-}$ and $\Det D^-(E)\cong \bigl(\Det D^+(E)
\bigr)\inv $ for any complex vector bundle~$E$ to rewrite the second factor
in~\thetag{4.2} as a pfaffian:
  $$ \Det D^+_{\alpha ,g}(K\inv _g) \cong  \Pf D^+_{\alpha ,g}(K\inv _g
     \oplus \Kbig). \tag{4.10} $$
In this expression $K\inv _g \oplus \Kbig$ refers to a rank two {\it real\/}
vector bundle.  Now apply the lemma to $[V]=\phi ^*[TY] - [K_g\inv \oplus
\Kbig ]$.  Assuming $Y^{10}$ to be oriented and spin, this has rank~8 and
vanishing~$w_1$ and~$w_2$.

        \demo{Proof of \theprotag{4.8} {Lemma}}
 A $KO$-theory version of the index theorem for flat bundles~\cite{APS2}
gives a topological formula for this difference of $\xi /2$-invariants
(see~\thetag{3.3}).  Let $[V]\in KO(P)$ denote the $KO$-class of~$V$.  The
difference of spin structures is a flat real bundle, and gives rise to a
class $[\ah-\bh]\in KO\inv (P;\qz.)$.  Let $\pi _!\:KO\inv (P;\qz)\to \qz$ be
the direct image map defined by the spin structure~$\bh$.  Then the flat
index theorem asserts
  $$ \frac{\xi _{\ah\phantom{\bh}}}2 - \frac{\xi _{\bh}}2 \equiv \pi
     _!\bigl([V]\cdot [\ah-\bh]\bigr)\pmod 1. \tag{4.11} $$
In fact, we will show that
  $$ [V]\cdot [\ah-\bh] = 0 \qquad \text{in $KO\inv (P;\qz)$}. \tag{4.12} $$
First, an element $[V]\in KO(P)$ is determined by $\rank([V])$, $w_1([V])$,
and~$w_2([V])$.  (An element of~$KO(P)$ is a homotopy class of maps
$P\to\ZZ\times BO$, and if the rank and first two Stiefel-Whitney classes are
trivial the map lands in~$BSpin$, which has trivial 3-skeleton.)  So with our
hypotheses $[V]=8k$ for some~$k\in \ZZ$.  Next, the difference of spin
structures is given by a homotopy class of maps $P\to \RP^{\infty}$, and
since $P$~is 3-dimensional by a map $P\to \RP^4$.  The rational (reduced)
$KO$~groups of~$\RP^4$ vanish, and the reduced
group~$\widetilde{KO}(\RP^4)\cong \ZZ/8\ZZ$, as was computed by
Adams~\cite{A}.  Hence $KO\inv (\RP^4;\qz)\cong \widetilde{KO} (\RP^4)\cong
\ZZ/8\ZZ$.  It follows that $8[\ah-\bh]=0$, whence \thetag{4.12}.
        \enddemo

        \demo{Proof of \theprotag{4.7} {Theorem}}
 We must show that for any family $\pi \:X\to Z$ of spin surfaces with the
usual geometric data, the line bundle
  $$ \scrL = (\Det D^+)^5 \otimes  (\Det\dbar)^{-5} \otimes \Det\dbar(K\inv )
     \otimes \bigl(\Det D^+(K\inv ) \bigr)\inv        $$
over~$Z$ is trivial.  (For readability, we omit the metric~$g$ and spin
structure~$\alpha $ from the notation in this proof.)  That $\scrL$~is
topologically trivial follows from a cohomological computation, thanks to~
\thetag{3.7}.  Namely, we can write~$\scrL$ as the determinant line bundle of
the Dirac operator~$D^+$ coupled to the virtual complex vector bundle
  $$ [W] = [5 - 5K^{-1/2} + K^{-3/2} - K^{-1}]\in K(X).   $$
Let $x=c_1(K^{-1/2})\in H^2(X)$.  Then
  $$ p_1\bigl([W] \bigr) = -5x^2 + 9x^2 -4x^2 =0, \tag{4.13} $$
and so $c_1(\scrL)=\pi _*p_1(W)=0$.  The curvature of the natural connection
on~$\scrL$ also vanishes, since the curvature~\thetag{3.1} is computed by a
combination of differential forms with the same coefficients as
in~\thetag{4.13}.

It remains to show that the holonomy is trivial.  For this we rewrite~$\scrL$
(canonically) as a product of pfaffian line bundles of Dirac operators
coupled to {\it real\/} vector bundles.  (For the last factor,
see~\thetag{4.10}):
  $$ \multline
      \scrL = \bigl(\Pf D^+ \bigr)^{10} \otimes \bigl(\Pf D^+(K^{-1/2} \oplus
     \Kbar^{-1/2}) \bigr)^{-5} \\
      \otimes \bigl(\Pf D^+(K^{-3/2}\oplus \Kbar^{-3/2}) \bigr) \otimes
     \bigl(\Pf D^+(K\inv \oplus \Kbar\inv ) \bigr)\inv .\endmultline
     \tag{4.14} $$
Consider a family of surfaces $P\to S^1$ fibered over the circle.  The
holonomy of~$\scrL$ around~$\cir$ is the adiabatic limit of the product of
$\tau^{-1/2}$-invariants of the Dirac operator~$D_P$ on~$P$ coupled to the
real vector bundles indicated in~\thetag{4.14}.  These bundles are associated
to the relative tangent bundle of the fibering $P\to\cir$.  We rewrite these
coupled Dirac operators as operators which make sense on {\it any\/} spin
3-manifold.  (Below we write~$P=\partial W$ for a spin 4-manifold~$W$ and we
want to extend the operators over~$W$, which may not fiber over the circle.)
For example, since $TP\otimes \CC\cong K\inv \oplus \Kbar\inv \oplus
1_{\CC}$, the operators which appear in the first and last factors
in~\thetag{4.14} may be combined as
  $$ D_P(11-TP), \tag{4.15} $$
where $11$~is the trivial real bundle of rank~11.  (What we really compute is
the $11^{\text{th}}$~power of the $\tau ^{-1/2}$-invariant of~$D_P$ divided by
the $\tau ^{-1/2}$-invariant of~$D_P(TP)$, but it is convenient to use virtual
bundles as a shorthand for this.)  For the second factor in~\thetag{4.14} we
rewrite $D_P(K^{-1/2} \oplus \Kbar^{-1/2})$ on the fibered manifold~$P\to
S^1$ as the operator\footnote{The notation~``$B^{\text{ev}}$'' is taken
from~\cite{APS1}.}
  $$ \Bev = (-1)^p(*d - d*)  $$
acting on $\Omega ^0(P)\oplus \Omega ^2(P)$, where $p=0$ on~$\Omega ^0$ and
$p=1$ on~$\Omega ^2$.  This operator is ``half'' the boundary of the four
dimensional signature operator, as explained in~\cite{APS1}.  An important
point for us is that its $\xi $-invariant is well-defined as a real number,
since the kernel of~$\Bev$ has a cohomological interpretation so has constant
dimension in families.  Hence $\xi /2\pmod1$ is well-defined.  For the third
term of~\thetag{4.14} we rewrite $D_P(K^{-3/2}\oplus \Kbar^{-3/2})$ on the
fibered manifold as
  $$ \Bev(TP) - 2\Bev.  $$
Thus the operators which appear in the second and third factors
in~\thetag{4.14} may be combined as
  $$ \Bev(TP-7). \tag{4.16} $$

Now we use the fact that any spin 3-manifold~$P$ bounds a spin
4-manifold~$W$.  Then by the Atiyah-Patodi-Singer index theorem for manifolds
with boundary~\cite{APS1}, the $\tau ^{-1/2}$-invariant is the exponential of
a curvature integral over~$W$.  The particular curvature polynomial is
determined from~\thetag{4.15} and~\thetag{4.16} to be the degree four part of
a form we denote schematically by
  $$ -\frac 12\Ahat(W)\bigl(12-\ch(TW)\bigr) - \frac 14L(W)\bigl(\ch(TW)-8
     \bigr). \tag{4.17} $$
Note that $TW\res P\cong TP\oplus 1$, which explains the~``12'' and~``8''
in~\thetag{4.17}.  Also, $L(W)$~is the {\it unstable\/} characteristic class
based on the formal expression $\frac x{\tanh x/2}$, as explained
in~\cite{AS1,p.~577}.  Expressed in terms of the first Pontrjagin
polynomial~$p_1$ in the curvature, the degree four part of~\thetag{4.17} is
the coefficient of~$p_1$ in
  $$ -\frac 12(1-\frac{p_1}{24})(8-p_1) - \frac 14(4+\frac{p_1}{3})(p_1-4),
      $$
which vanishes.  Thus the holonomy of~$\scrL$ is trivial.
        \enddemo

 \head
 \S{5} Open Superstrings and D-Branes
 \endhead
 \comment
 lasteqno 5@ 27
 \endcomment

\noindent{\it Statement of Results}
\smallskip

Fix a Riemannian manifold~$Y$, which we assume to be oriented, spin, and of
dimension~10.  As before, this is the background in which the string
propagates; now we want to add a D-brane.  Thus let $Q\subset Y$ be an
oriented submanifold, which we need not assume to be connected.  The oriented
(not necessarily connected) surface~$\Sigma $ is now permitted to have a (not
necessarily connected) boundary, and we require the boundary of~$\Sigma $ to
map to~$Q$.  In other words, the space of bosons is
  $$ B = \Met(\Sigma )\times \Map\bigl((\Sigma ,\bS),(Y,Q)\bigr).
     $$
Asking that $\bS$~map to $Q$ imposes a mixture of Dirichlet and Neumann
boundary conditions.  To define the determinant
lines~\thetag{4.2} and~\thetag{4.3}, we
need to impose boundary conditions on the fermions; the desired
boundary conditions are local boundary conditions that are determined by
supersymmetry.  These boundary conditions mix right- and left-handed spinor
fields, so it is only the tensor product of the lines ${\Cal L}^+$ and
${\Cal L}^-$ which makes sense if
the boundary is nonempty.

To describe precisely the desired Dirac operators on $\Sigma$,
we first recall that a spin structure~$\alpha $
on~$\Sigma $ induces a spin structure on~$\bS$.  To see this, fix a
metric~$g$ on~$\Sigma $ and consider the principal $SO_2$-bundle of oriented
orthonormal frames $SO(\Sigma )\to\Sigma $.  Its restriction to~$\bS$ is
canonically trivialized by the oriented orthonormal frame whose first element
is the outward pointing unit normal.  A spin structure~$\alpha $ induces a
double cover $\Spin(\Sigma )\to SO(\Sigma )$, and the inverse image of the
trivialization at the boundary is a spin structure on the boundary.  That
inverse image is a double cover of~$\bS$.  There are two possibilities on
each component of~$\bS$.  If the double cover is connected, then we say that
the spin structure on that component is trivial, since it bounds a spin
structure on a disk.  If the double cover is not connected, then we say that
the spin structure on that component is nontrivial.  The spin structure on
the entire boundary~$\bS$ is constrained by the fact that it is the boundary
of a spin structure on~$\Sigma $.  As for the complex line
bundle~$K^{1/2}_{\alpha ,g}$, it follows from this discussion that its
restriction to~$\bS$ has a canonical real structure, i.e.,
  $$ K^{1/2}_{\alpha ,g} \cong \overline{K^{1/2}_{\alpha ,g}}\qquad
     \text{on~$\bS$}.  \tag{5.1} $$
The underlying real bundle determines the spin structure on~$\bS$.

As for closed strings we allow different spin structures on the right- and
left-movers.  Thus we consider pairs of spin structures~$\alpha ,\beta $ such
that the induced spin structures on~$\bS$ are isomorphic.  Now an isomorphism
of spin structures on~$\bS$ is only determined up to a sign on each component
of~$\bS$, and we need to fix that sign.  (The overall sign is irrelevant.)
Thus part of the data we need is an isomorphism
  $$ \theta \: K^{1/2}_{\alpha }\res{\bS}\longrightarrow
     K^{1/2}_{\beta}\res{\bS}. \tag{5.2} $$
Once the isomorphism is chosen for some metric it is determined for all
metrics, so \thetag{5.2}~is a discrete topological choice.  Thus the
topological data is a triple~$(\alpha ,\beta ,\theta )$.

Let $D$~denote the total Dirac operator, which is the sum of the two chiral
operators~\thetag{4.1}.  For a map~$\phi \:\Sigma \to Y$ with $\phi
(\bS)\subset Q$ and a metric~$g$ on~$\Sigma $, consider the total Dirac
operator $D_{\alpha, \beta ,\theta ,g}(\phi ^*TY)$, where we use the spin
structure~$\alpha $ on right-handed spinors and the spin structure~$\beta $
on left-handed spinor fields.  The isomorphism~$\theta $ does not enter into
the definition of the Dirac operator, but is included in the notation since
it does enter into the boundary condition~\thetag{5.3} below.  A spinor
field~$\psi $ with values in~$\phi ^*TY$ decomposes as $\psi =\psi ^++\psi
^-$ according to the chiral decomposition of spinors on~$\Sigma $.  The
restriction~$\partial \psi $ of~$\psi $ to~$\bS$ takes values in $\phi
^*(TY\res Q)$.  Now
  $$ TY\res Q \cong TQ \oplus \nu ,  $$
where $\nu $~is the normal bundle to~$Q$ in~$Y$.  Thus at the boundary we can
write
  $$ \partial \psi ^{\pm} = \lambda ^\pm_Q + \lambda ^\pm_\nu .  $$
The boundary condition for the operator~$D_{\alpha , \beta ,\theta ,g}(\phi
^*TY)$ is then
  $$ D_{\alpha ,\beta ,\theta ,g}(\phi ^*TY):\qquad\aligned
      \theta (\lambda ^{+}_Q) &= \hphantom{-}\lambda ^-_Q \\
      \theta (\lambda ^{+}_\nu ) &= -\lambda ^-_\nu \endaligned \tag{5.3} $$
The operator $D_{\alpha , \beta ,\theta ,g}(\phi ^*TY)$ with these boundary
conditions is complex skew-adjoint.

There are two other operators---acting on the ghost fields---which enter.
They appear in the determinant line of interest:
  $$ \multline
      L_{\alpha,\beta ,\theta } (g,\phi ) = \Pf D_{\alpha,\beta ,\theta
     ,g}(\phi ^*TY) \otimes \bigl[ \Det D_{\alpha ,\beta ,\theta ,g}(K_g\inv
     \oplus \Kbig )\bigr]^{-1} \\
      \otimes \Det \bigl(\dbar_g(K_g\inv ) \oplus \partial_g
     (\Kbig)\bigr).\endmultline  $$
For each of the last two factors we ask that the two fields which appear be
equal on~$\partial \Sigma $.  This boundary condition makes sense, since
on~$\partial \Sigma $ there is a natural real structure on the inverse
canonical bundle~$K_g\inv $.  Note that in the second factor the boundary
condition involves the isomorphism~$\theta $, as in~\thetag{5.3}.  The field
in the third factor is a complexified tangent vector to the surface, and the
boundary condition is the complexification of the condition that a real
tangent vector be tangent to the boundary.  Thus the domain of the operator
is the complexification of the group of infinitesimal diffeomorphisms.  The
boundary condition in the second factor is the odd analog for
superdiffeomorphisms.

We study~$L_{\alpha ,\beta ,\theta }$ as a line bundle over~$Z$ for
families~\thetag{4.4}, where now $X$~has a boundary which fibers over~$Z$
with typical fiber~$\bS$ and $\phi (\partial X)\subset Q$:
  $$ \CD
     \partial X @>\partial \phi >> Q \\
     @VV\partial \pi V\\
     Z \endCD  \tag{5.4} $$

Our first result computes the topology---that is, the first Chern class---of
the line bundle~$L_{\alpha ,\beta ,\theta }$.

        \proclaim{\protag{5.5} {Theorem}}
 The isomorphism class of the complex line bundle~$L_{\alpha ,\beta ,\theta
}$ equals $(\partial \pi) _*(\partial \phi) ^*W_3(\nu )$
in~$H^2(Z;\ZZ)$, where $W_3$~is the third Stiefel-Whitney class.  In fact,
this line bundle is the complexification of a real line bundle, and the
isomorphism class in~$H^1(Z;\ZZ/2\ZZ)$ of the underlying real line bundle
equals $(\partial \pi) _*(\partial \phi) ^*w_2(\nu )$.
        \endproclaim

\flushpar
 Here $(\partial \pi )_*$~denotes integration along the fibers of~$\partial
\pi $.  The first statement follows from the second by applying the Bockstein
homomorphism, which commutes with pushforward and pullback.

As explained at the beginning of~\S{3}, for the application to anomalies the
topological isomorphism class is not fine enough.  One needs also to compute
the isomorphism class of the canonical connection---its curvature and
holonomy.  The theory of the Quillen metric and canonical connection on the
determinant line bundle only exists in the literature for families of closed
manifolds~\cite{BF}, or families of manifolds with boundary and {\it
global\/} boundary conditions of Atiyah-Patodi-Singer type~\cite{P}.  In our
problem we have a family of manifolds with boundary and {\it local\/}
boundary conditions.  As we will see below, by gluing we identify the index
problem on surfaces with boundary with an index problem on the double
manifold, and then we can apply the usual geometric theory of determinant
line bundles on closed manifolds.  In~\S{6} we describe a general class of
Dirac operators with local boundary conditions and the associated doubling.

Let $X^d\to Z$~be the family of doubled surfaces, and $\gamma =\gamma (\alpha
,\beta ,\theta )$ the spin structure on the double obtained by gluing~$\alpha
,\beta $ using~$\theta $.  As part of the proof of \theprotag{5.5} {Theorem}
we identify $L_{\alpha ,\beta ,\theta }$ with a twisted version~$\Lt_\gamma
^+$ of~\thetag{4.2}.  Of course, the topology of~$\Lt_\gamma ^+$ is
determined by~$W_3(\nu )$, as in \theprotag{5.5} {Theorem}, but our main
result asserts that the holonomy of its natural connection is given by the
second Stiefel-Whitney class~$w_2(\nu )$.

        \proclaim{\protag{5.6} {Theorem}}
 The canonical connection on~$\Lt_\gamma ^+$ is flat.  Consider a family of
surfaces with boundary $\pi\:X\to S^1$ and the associated family of doubles
$\pi ^d\:X^d\to S^1$.  Then the holonomy of the canonical connection around
this loop is~$\pm 1$ with the sign given by $(\partial \phi )^*w_2(\nu
)[\bX]$.  \rom(See~\thetag{5.4} for the notation.\rom)
        \endproclaim

\bigskip\noindent{\it Proofs}
\smallskip\nobreak
The proof of \theprotag{5.5} {Theorem} proceeds in two main steps.  Define an
operator $D'_{\alpha ,\beta ,\theta ,g}(\phi ^*TY)$ which differs from
$D_{\alpha ,\beta ,\theta ,g}(\phi ^*TY)$ only by a sign in the boundary
condition:
  $$ D'_{\alpha ,\beta ,\theta ,g}(\phi ^*TY):\qquad\aligned
      \theta (\lambda ^{+}_Q) &= \lambda ^-_Q \\
      \theta (\lambda ^{+}_\nu ) &= \lambda ^-_\nu \endaligned \tag{5.7} $$
The operator $D'_{\alpha ,\beta ,\theta ,g}(\phi ^*TY)$ is also complex skew
adjoint.  Then set
  $$ \multline L'_{\alpha,\beta ,\theta } (g,\phi ) = \Pf D'_{\alpha,\beta
     ,\theta ,g}(\phi ^*TY) \otimes \bigl[ \Det D_{\alpha ,\beta ,\theta
     ,g}(K_g\inv \oplus \Kbig )\bigr]^{-1} \\
      \otimes \Det \bigl(\dbar_g(K_g\inv ) \oplus \partial_g
     (\Kbig)\bigr).\endmultline  $$
We first show the following.

        \proclaim{\protag{5.8} {Proposition}}
 The ratio $L_{\alpha ,\beta ,\theta } \otimes (L'_{\alpha ,\beta ,\theta
})\inv $ is the complexification of a real line bundle whose isomorphism
class in $H^1(Z;\ZZ/2\ZZ)$ is $(\partial \pi) _*(\partial \phi) ^*w_2(\nu
)$.
        \endproclaim

\flushpar
 In the process of proving~\theprotag{5.8} {Proposition} we identify
$L'_{\alpha ,\beta ,\theta }$ with the line bundle~$L_\gamma ^+$
(see~\thetag{4.2}) on the double~$\Xd$, where as above $\gamma =\gamma
(\alpha ,\beta ,\theta )$ is the spin structure on the double obtained by
gluing~$\alpha ,\beta $ using~$\theta $.  We also identify~$L_{\alpha ,\beta
,\theta }$ with a twisted version~$\Lt_\gamma ^+$.  The first step towards
proving~\theprotag{5.6} {Theorem} is a geometric version of \theprotag{5.8}
{Proposition}.

        \proclaim{\protag{5.9} {Proposition}}
 The canonical connection on the ratio~$\Lt_\gamma ^+\otimes (L^+_\gamma
)\inv $ is flat, and the holonomy for any family of {\it closed\/} surfaces
$\pi :\scrX^d\to S^1$ is~$\pm 1$ with the sign given by $(\partial \phi
)^*w_2(\nu )[\bsX]$.
        \endproclaim

The second step in the proofs of \theprotag{5.5} {Theorem} and
\theprotag{5.6} {Theorem} is the following.

        \proclaim{\protag{5.10} {Proposition}}
 $L^+_\gamma $~is trivializable (as a bundle with metric and connection) over
a family of doubled surfaces.
        \endproclaim

\flushpar
 We prove \theprotag{5.10} {Proposition} directly only for the symmetric
case~$\alpha =\beta $; the result for arbitrary~$\alpha ,\beta $ then follows
from \theprotag{4.5} {Theorem}.

We give three proofs of \theprotag{5.8} {Proposition}.  Notice that the line
bundle in question is
  $$ \scrL_{\alpha ,\beta ,\theta }:= (\Pf D\mstrut _{\abt}) \otimes (\Pf
     D'_{\abt})\inv . \tag{5.11} $$

        \demo{First Proof of \theprotag{5.8} {Proposition}}
 A general result~\cite{BW, Theorem 21.2} for elliptic boundary-value
problems (the ``Agranovi\v c-Dynin formula'') describes the dependence of the
index on {\it local\/} boundary conditions.  The version we need is for
families of complex skew-adjoint Dirac operators, and it follows
from~\cite{N,Theorem 6.2}.  It computes the difference of the $KO$-indices
of~$D_{\abt}$ and~$D'_{\abt}$ as the index of a family of operators on the
boundary family~\thetag{5.4}:
  $$ \ind^{X/Z}(D_{\abt}) - \ind^{X/Z}(D'_{\abt}) = \ind^{\partial X/Z}
     (RPR^{\prime *})\quad \in KO^{-2}(Z). \tag{5.12} $$
We must explain the operators $P,R,R'$ on the right hand side.

First, $P$~is a zeroth order pseudodifferential operator, the {\it Calder\'on
projector\/}.  We only need its principal symbol.  For that, recall that the
spinor fields on the boundary are sections of
  $$ \bigl(K^{1/2}_\alpha \oplus \Kbar_\beta ^{1/2}\bigr) \otimes (\partial
     \phi )^*(TQ\oplus \nu ). \tag{5.13} $$
Also, using~\thetag{5.1} and~\thetag{5.2} we identify the two spinor bundles
as a single real line bundle~$S$.  For simplicity introduce the notation
  $$ \aligned
      F &:= (\partial \phi )^*TQ \\
      F' &:= (\partial \phi )^*\nu .\endaligned \tag{5.14} $$
Then we rewrite~\thetag{5.13} as the real bundle
  $$ (S\oplus S)\otimes (F\oplus F') \cong \bigl(S\otimes (F\oplus F')\bigr)
     \;\oplus \; \bigl(S\otimes (F\oplus F')\bigr). \tag{5.15} $$
Let $\Dirac$~be the standard Dirac operator on the boundary circle.  Then the
operator which acts on~\thetag{5.14}---the ``boundary operator'' associated
to~$D_{\abt}$ (see~\cite{BW,Theorem 12.4})---is $\left(\smallmatrix
\sDirac\\&&-\sDirac \endsmallmatrix\right)$ relative to the
decomposition~\thetag{5.15}.  The principal symbol of this operator is
invertible and self-adjoint.  The Cald\'eron projector~$P$ has a principal
symbol~$\sigma (P)$ which is the projection onto the sum of the eigenspaces
with positive eigenvalues.  Thus for a nonzero cotangent vector~$\xi $ on the
boundary,
  $$ \sigma (P)(\xi ) = \cases \left(\smallmatrix 1\\&&0
     \endsmallmatrix\right) ,&\xi >0;\\\left(\smallmatrix 0\\&&1
     \endsmallmatrix\right),&\xi <0,\endcases \tag{5.16} $$
where we use the orientation of the boundary to give meaning to the sign
of~$\xi $.

The operators $R,R'$ in~\thetag{5.12} are vector bundle maps
  $$ R,R'\:\bigl(S\otimes (F\oplus F') \bigr) \;\oplus\; \bigl(S\otimes
     (F\oplus F') \bigr) \longrightarrow S\otimes (F\oplus F')  \tag{5.17} $$
whose kernel consists of spinor fields which satisfy the local boundary
conditions~\thetag{5.3}, \thetag{5.7}.  Relative to the decompositions shown
in~\thetag{5.17}, we write these operators as the matrices
  $$ \aligned
      R &= \pmatrix 1\oplus 1&-1\oplus 1  \endpmatrix, \\
      R' &= \pmatrix 1&-1  \endpmatrix .\endaligned \tag{5.18} $$
Since $R,R'$~are vector bundle maps, they are equal to their principal
symbols.

Now the principal symbol of the operator~$RPR^{\prime *}$ which appears
in~\thetag{5.12} is easily computed from~\thetag{5.16} and~\thetag{5.18}:
  $$ \sigma (RPR^{\prime*})(\xi ) = \cases \left(\smallmatrix 1\oplus
     1&&-1\oplus 1 \endsmallmatrix\right)\left(\smallmatrix 1\\&&0
     \endsmallmatrix\right)\left(\smallmatrix \hphantom{-}1\\-1
     \endsmallmatrix\right)=\left(\smallmatrix 1 \endsmallmatrix\right) ,&\xi
     >0;\\
      \left(\smallmatrix 1\oplus 1&&-1\oplus 1
     \endsmallmatrix\right)\left(\smallmatrix 0\\&&1
     \endsmallmatrix\right)\left(\smallmatrix \hphantom{-}1\\-1
     \endsmallmatrix\right)=\left(\smallmatrix 1 \oplus
     -1\endsmallmatrix\right) ,&\xi <0.\endcases \tag{5.19} $$
These matrices act on $S\otimes (F\oplus F')$.  Observe that \thetag{5.19}
{}~is also the symbol of the family of operators
  $$ \id_{S\otimes F} \;\oplus\;  \Dirac(F')  $$
on $\partial X\to Z$.  Since the index only depends on the symbol, and since
the index of the identity operator is trivial, we see that \thetag{5.12}
reduces to
  $$ \ind^{X/Z}(D_{\abt}) - \ind^{X/Z}(D'_{\abt}) = \ind^{\partial
     X/Z} \bigl(\Dirac(F')\bigr).  $$
The index we are computing lies in~$KO^{-2}(Z)$; it is the index of a family
of {\it complex\/} skew-adjoint operators.  But $\Dirac(F')$ is a family of
{\it real\/} skew-adjoint operators, so its index lies naturally in~$KO\inv
(Z)$.  The index we seek in~$KO^{-2}(Z)$ is thus the image of $\ind^{\partial
X/Z}\Dirac(F')$ under the natural map
  $$ KO\inv (Z) @>[\eta -1]>>KO\inv (\pt)\otimes  KO\inv (Z)  \cong
     KO^{-2}(Z), \tag{5.20} $$
where $\eta \in KO(S^1)$ is the M\"obius bundle.  We are interested in the
complex pfaffian line bundle, which in this situation is the complexification
of the real pfaffian line bundle:
  $$ \CD
      KO\inv (Z) @>\thetag{5.20}>> KO^{-2}(Z) \\
      @VV\Pf_{\RR} V @VV\Pf_{\CC} V\\
      H^1(Z;\ZZ/2\ZZ) @>\beta >> H^2(Z;\ZZ)\endCD \tag{5.21} $$
Here $\beta $~is the Bockstein map.  So we are reduced to demonstrating a
well-known result:  for any fibering $\rho \:W\to Z$ of spin 1-manifolds and
any oriented real vector bundle $F'\to W$,
  $$ \Pf_{\RR} \Dirac^{W/Z}(F') = \rho _* w_2(F'). \tag{5.22} $$

To check~\thetag{5.22} it suffices to take~$Z=\cir$, so $W=\cir\times \cir$.
Endow~$Z$ with the bounding spin structure; then the left hand side
of~\thetag{5.22}, evaluated on the fundamental class of~$Z$, is the image
of~$\ind^{W/Z}\Dirac(F')$ under
  $$ \pi _!^Z\:KO\inv (\cir)\longrightarrow KO^{-2}(\pt)\cong
     \ZZ/2\ZZ. \tag{5.23} $$
By the multiplicative property of the direct image---essentially the
Thom isomorphism in $KO$-theory---this is $\pi _!^W\bigl([F'] \bigr)$ for
  $$ \pi _!^W\:KO(W)\longrightarrow KO^{-2}(\pt)\cong \zt  $$
the direct image map.  (By the index theorem~\cite{AS2} this is the mod~2
index of the Dirac operator on~$W$ coupled to the real vector bundle~$F'$.)
Since $W$~bounds a spin manifold, we have $\pi _!^W\bigl([1] \bigr)=0$, and
so
  $$ \pi _!^W\bigl([F'] \bigr) = \pi _!^W\bigl([F'-\dim F']
     \bigr).  $$
Now $x = [F'-\dim F']\in \widetilde{KO}(W)$ satisfies $w_1(x) = w_{1}(F')=0$,
so is determined by $w_2(x) = w_2(F')$.  We can arrange the support of~$x$ to
be contained in a disk in~$W$, and by excision it follows that $\pi _!^W(x) =
\pi _!^{S^2}(y)$ for $y\in \widetilde{KO}(S^2)$ with $w_2(y)[S^2]=
w_2(x)[W]$.  Finally, the direct image map
  $$ \pi _!^{S^2}\:\widetilde{KO}(S^2)\longrightarrow KO^{-2}(\pt)\cong \zt
     \tag{5.24} $$
is an isomorphism, and the generator of~$\widetilde{KO}(S^2)$ is a bundle
with $w_2\not= 0$.
        \enddemo

For the other proofs of \theprotag{5.8} {Proposition} we introduce the
fibered double $\Xd\to Z$, which is a family of closed oriented surfaces
obtained by gluing $X\cup -X$ along the boundary~$\bX$.  Also, the spin
structure~$\alpha $ on~$X/Z$ and $\beta $ on~$-X/Z$ glue to a spin
structure~$\gamma $ on~$\Xd/Z$ via the isomorphism~$\theta $ in~\thetag{5.2}.
A right-handed spinor field on~$\Xd_\gamma /Z$ can be identified with a
pair~$(\psi ^+,\psi ^-)$: $\psi ^+$~is a right-handed spinor field
on~$X_\alpha $, $\psi ^-$~is a left-handed spinor field on~$X_\beta $, and
$\psi ^+=\psi ^-$ on~$\bX$ using~$\theta $.  So the operator $D'_{\alpha
,\beta ,\theta ,g}(\phi ^*TY)$ (see~\thetag{5.7}) may be
identified\footnote{The identification of the indices under gluing is
discussed in a similar situation in~\cite{F3,\S2}.  The gluing indicated here
only gives continuous spinor fields on the double; it is nicer to glue along
an open cylinder near the boundary, as in~\cite{DF,\S IV}.  We generalize and
give more details in the second part of section~6.} with the chiral Dirac
operator~$D^+_{\gamma ,g}(\phi ^*TY)$ on the double.  Note that $X^d/Z$~has
an orientation-reversing involution, but it does not lift to the spin bundle
unless~$\alpha =\beta $ and $\theta $~is the identity.

Next, we rewrite the twisted boundary conditions~\thetag{5.3} on the double,
for simplicity on a single surface~$\Sigma $.  Then in the double~$\Sd$ there
is a collar neighborhood $I\times \partial \Sigma \subset \Sd$ of~$\partial
\Sigma $, where $I\simeq(-1,1)$.  Over that neighborhood we have a splitting
$\phi ^*TY\cong F\oplus F'$, using the notation of~\thetag{5.14}.  Let $\eta
$~be the real line bundle on~$\Sd$ obtained by identifying the trivial real
line bundle on~$\Sigma $ with the trivial real line bundle on~$-\Sigma $ via
the isomorphism~$-1$ on the collar region.  Then $\eta $~is canonically
trivial away from the collar, and there is a real vector bundle $E\to\Sd$
which is canonically~$\phi ^*TY$ away from the collar and $F\oplus (\eta
\otimes F')$ on the collar.  We identify $D_{\alpha ,\beta ,\theta ,g}(\phi
^*TY)$ on~$\Sigma $ with $D^+_{\gamma ,g}(E)$ on~$\Sd$; the line bundle~$\eta
$ incorporates the sign in~\thetag{5.3}.  Therefore, the ratio~\thetag{5.11}
for a family $X\to Z$ of surfaces with boundary is
  $$ \Pf _{\CC}D^+_\gamma \bigl((\eta -1)\otimes F' \bigr) \tag{5.25} $$
for the family of chiral Dirac operators on the relative double $\Xd\to Z$.
Note that $\eta -1$~is supported in the collar region~$I\times \partial X$
and is pulled back from the first factor.  We identify
  $$ [\eta -1]\in KO(I,\partial I)  $$
as the (Hopf bundle) generator.  Also, up to isomorphism $F'$~is pulled back
from the second factor~$\bX$:
  $$ [F'] = \bigl[(\partial \phi )^*\nu \bigr] \in  KO(\partial
     X).  $$

        \demo{Second proof of \theprotag{5.8} {Proposition}}
 By the excision property of the index, \thetag{5.25}~is
  $$ \Pf_{\CC} \pi _!^{(I,\partial I)\times \bX/Z}\bigl((\eta -1)\otimes [F']
     \bigr),  $$
where
  $$ \pi _!^{(I,\partial I)\times \bX/Z}\:KO\bigl((I,\partial I)\times \bX
     \bigr) \longrightarrow KO^{-2}(Z)  $$
is the direct image map.  This factors as
  $$ \pi _!^{(I,\partial I)\times \bX/Z}\bigl((\eta -1)\otimes [F']
     \bigr) = \pi _!^{(I,\partial I)}\bigl([\eta -1] \bigr) \cdot \pi
     _!^{\bX/Z}\bigl([F'] \bigr) \in KO^{-1}(\pt)\otimes
     KO^{-1}(Z). \tag{5.26} $$
Multiplication by~$[\eta -1]$ is~\thetag{5.20}, and the computation of $\pi
_!^{\bX/Z}\bigl([F'] \bigr)$ proceeds as in the first proof.
(See~\thetag{5.22} and the argument which follows.)
        \enddemo

        \demo{Third proof of \theprotag{5.8} {Proposition}}
 Assume first that $\rank F'$~is even.  Write $x=(\eta -1)\otimes F'$.  Then
from the Whitney sum formula we compute the Stiefel-Whitney classes of the
virtual bundle~$x$:
  $$ \aligned
      &w_1(x)=w_2(x)=w_3(x)=0, \\
      &w_4(x)=w_1(\eta )\smile w_3(F'). \endaligned  $$
Thus $x$~is orientable and spinable, so by~\thetag{3.6} the isomorphism class
of the pfaffian line bundle is a pushforward in cohomology:
  $$ c_1\Pf D^+_\gamma (x) = \pi _*^{\Xd/Z}\bigl(\lambda (x) \bigr),
      $$
where
  $$ \pi _*^{\Xd/Z}\:H^4(\Xd)\longrightarrow H^2(Z)  $$
and $\lambda $~is the characteristic class of a spin bundle with~$2\lambda
=p_1$.  By excision, we compute on the tubular neighborhood $(I,\partial
I)\times \bX$ and extend to~$\Xd$.  We claim
  $$ \lambda (x) = \beta \bigl(w_1(\eta )\smile w_2(F') \bigr)\in
     H^4\bigl((I,\partial I)\times \bX \bigr); \tag{5.27} $$
then \theprotag{5.8} {Proposition} follows from~\thetag{5.21} and the fact
that the Bockstein~$\beta $ commutes with pushforward.  To prove the claim,
we work in the universal situation where the bundle $x=(\eta -1)\otimes
F'$ lives over~$\cir\times BSO$.  First, since $\eta \oplus \eta $~is
trivial, it follows that $\lambda (x)$~is torsion of order two.  Since all of
the torsion in~$H^{\bullet}(BSO)$ has order two (see~\cite{BH,\S30}) it
follows that $\lambda (x)$~is determined by its image in real cohomology,
which vanishes, and its image in $\zt$~cohomology, which is $w_4(x)=w_1(\eta
)\smile w_3(F')$.  The integral class in~\thetag{5.27} is also torsion of
order two and reduces mod~2 to $w_1(\eta )\smile w_3(F')$.

If $\rank F'$~is odd, we replace~$F'$ with $F'$ plus a trivial line bundle
and apply the previous.  So we are reduced to proving that $\Pf D^+_\gamma
(\eta -1)$ is trivial.  This follows from excision and the
factorization~\thetag{5.26}.
        \enddemo

We now prove \theprotag{5.9} {Proposition}.  Notice that the ratio of bundles
which appears in this proposition is~\thetag{5.25}.

        \demo{Proof of \theprotag{5.9} {Proposition}}
 The curvature~\thetag{3.1} vanishes since virtual bundle~$(\eta -1)\otimes
F'$ has rank zero and is flat.  To compute the holonomy~\thetag{3.2} around a
family of closed manifolds $\pi \:\scrX\to\cir$, we endow the base~$S^1$ with
the bounding spin structure and induce a spin structure on~$\scrX$; then the
holonomy is the ratio of the $\tau ^{-1/2}$-invariant of~$\eta \otimes F'\to
\scrX$ to the $\tau ^{-1/2}$-invariant of $F'\to\scrX$.  (There is no adiabatic
limit necessary since the connection is flat.)  By a real version of the flat
index theorem~\cite{APS2}, the difference of these bundles determines a class
$x\in KO\inv (\scrX;\zt)$, and the ratio of the holonomies is $\pi
_!^{\scrX}(x)\in KO^{-4}(\pt;\zt)\cong \zt$ (see~\thetag{3.3}).  As in previous
arguments we use excision to localize the computation to $(I,\partial
I)\times \bsX$, where we write~$x$ as the product of $[\eta -1]\in KO\inv
\bigl((I,\partial I);\zt \bigr)$ and $[F']\in KO(\bsX)$.  Thus
  $$ \pi _!^{\scrX}(x) = \pi _!^{(I,\partial I)}\bigl([\eta -1] \bigr)\cdot \pi
     _!^{\bsX}\bigl([F'] \bigr)\in \zt.  $$
The first factor is~1 and the second is computed in
\thetag{5.23}--\thetag{5.24} to be $(\partial \phi )^*w_2(\nu )[\bsX]$, as
desired.
        \enddemo

We proceed to the second step of the proof of \theprotag{5.5} {Theorem},
which is the proof of \theprotag{5.10} {Proposition}.  We demonstrate that
$L^+_\gamma $~is trivializable geometrically, that is, as a bundle with
metric and connection.

        \demo{Proof of \theprotag{5.10} {Proposition}}
 Consider first the case when~$\alpha =\beta $ and $\theta $~is the identity
map.  Then the orientation-reversing involution of the double manifold lifts
to the spin bundle.  In this case we claim that each of the three factors
in~\thetag{4.2} is trivializable.  First, consider a family of surfaces with
boundary $\pi \:X\to Z$ and the associated family of doubles $\pi ^d\:X^d\to
Z$.  The determinant and pfaffian bundles are flat since the
curvature~\thetag{3.1} is the integral over the fibers of~$\pi ^d$ of a
differential form which is invariant under the orientation-reversing
involution on the double, so vanishes.  Next, we investigate the holonomy.
Consider a family of surfaces with boundary $\pi \:X\to\cir$ and the
associated double $\pi ^d\:\Xd\to\cir$.  We endow~$\cir$ with the bounding
spin structure and lift to a spin structure on~$\Xd$.  The holonomy of the
first factor of~\thetag{4.2} is computed by a certain $\tau
^{-1/2}$-invariant and that of each of the last two factors by a $\tau
$-invariant.  Since $\Xd$~is the {\it spin\/} double of~$X$, we compute these
invariants using~\thetag{3.5}.  Thus in each case the invariant
is~$(-1)^{k\choose 2}$, where $k$ is the number of components of~$\bX$ with a
nontrivial index (or mod~2 index) of the appropriate boundary Dirac operator.
For the last factor the operator on any component of the boundary is the
boundary of a family of operators on the disk, so the index vanishes.  For
the first factor we note that the bundle $(\partial \phi )^*(TY)\to \bX$ is
trivializable, since it is oriented and spin, and now since the base~$\cir$
bounds a disk the mod~2 index vanishes.  A similar argument applies to the
second factor: rewrite the determinant line bundle as a pfaffian line bundle,
as in~\thetag{4.10}.

For general~$\alpha ,\beta ,\theta $ the result now follows from
\theprotag{4.5} {Theorem}.
        \enddemo

This completes the proofs of \theprotag{5.5} {Theorem} and \theprotag{5.6}
{Theorem}.

 \head
 \S{6} Additional Remarks
 \endhead
 \comment
 lasteqno 6@ 12
 \endcomment

\noindent{\it The $A$- and $B$-Fields}
\smallskip \nobreak
 Fields which are locally differential forms often have a nontrivial global
structure.  This was explained a bit in~\S1; here we add a few details.
Mathematical foundations for the low degree case needed here are developed
in~\cite{B}, though we do not use that language.\footnote{As remarked in the
introduction, a non-\v Cech mathematical theory adequate for all examples in
string theory and $M$-theory is lacking.}  For a related exposition,
see~\cite{DeF,\S6}.

We distinguish four types of $p$-form fields, two types with nontrivial
global structure together with their field strengths.  For $p$~small we can
say what they are in familiar geometric language:
$$
\vcenter{\offinterlineskip
\hrule
\halign{\vrule# &\quad\hfill #\hfill
  &\quad\vrule# &\quad\hfill #\hfill
  &\quad\vrule# &\quad\hfill #\hfill
  &\quad\vrule# &\quad\hfill #\hfill
  &\quad\vrule# &\quad\vrule# \cr
height6pt &\omit  &&\omit &&\omit &&\omit  &\cr
&notation&&$p$&&global description&&global $p$-form?&\cr
height6pt &\omit  &&\omit &&\omit &&\omit &\cr
\noalign{\hrule height 1.5pt depth 0pt}
height3pt &\omit   &&\omit &&\omit &&\omit &\cr
&$\Theta ^1$ &&$1$ &&connection on principal $\TT$-bundle&&no &\cr
height3pt &\omit &&\omit &&\omit &&\omit  &\cr
\noalign{\hrule}
height3pt &\omit  &&\omit &&\omit &&\omit &\cr
&$\Omega ^2_{\ZZ}$ &&$2$ &&curvature of connection&&yes &\cr
height3pt &\omit &&\omit &&\omit &&\omit  &\cr
\noalign{\hrule}
height3pt &\omit  &&\omit &&\omit &&\omit &\cr
&$\Gamma ^0$ &&$0$ &&section of principal $\TT$-bundle with connection&&no &\cr
height3pt &\omit &&\omit &&\omit &&\omit  &\cr
\noalign{\hrule}
height3pt &\omit  &&\omit &&\omit &&\omit &\cr
&$\Omega ^1$ &&$1$ &&covariant derivative of section&&yes &\cr
height3pt &\omit &&\omit &&\omit &&\omit  &\cr
\noalign{\hrule}
}
\hrule}
\tag{6.1} $$
\noindent
 A principal circle~($\TT$) bundle over a manifold~$M$ is a manifold~$P$ on
which $\TT$~acts freely with quotient~$M$.  The bundle~$P\to M$ is classified
topologically by its first Chern class in integral degree two cohomology.
Equivalently, we can view~$P$ as the set of unit vectors in a hermitian line
bundle over~$M$.  A connection is an imaginary 1-form on~$P$ which satisfies
some affine equations---it is not a differential form on the base~$M$.  The
notion that a 1-form in field theory is often such a connection is quite
familiar.  The curvature is a {\it closed\/} 2-form on the base~$M$ whose
periods are integer multiples of~$2\pi i$.  A section, or trivialization,
of~$P$ is a map $M\to P$ which splits the projection $P\to M$.  Equivalently,
it is a unit norm section of the associated hermitian line bundle.  In
general such sections exist locally; the Chern class of~$P$ is an obstruction
to global existence.  The covariant derivative of a section is the pullback
of the connection form to~$M$, a global 1-form which is not necessarily
closed.

The analog of the first two lines in lower degree may also be stated in
familiar terms:
$$
\vcenter{\offinterlineskip
\hrule
\halign{\vrule# &\quad\hfill #\hfill
  &\quad\vrule# &\quad\hfill #\hfill
  &\quad\vrule# &\quad\hfill #\hfill
  &\quad\vrule# &\quad\hfill #\hfill
  &\quad\vrule# &\quad\vrule# \cr
height6pt &\omit  &&\omit &&\omit &&\omit  &\cr
&notation&&$p$&&global description&&global $p$-form?&\cr
height6pt &\omit  &&\omit &&\omit &&\omit &\cr
\noalign{\hrule height 1.5pt depth 0pt}
height3pt &\omit   &&\omit &&\omit &&\omit &\cr
&$\Theta ^0$ &&$0$ &&map to $\TT$&&no &\cr
height3pt &\omit &&\omit &&\omit &&\omit  &\cr
\noalign{\hrule}
height3pt &\omit  &&\omit &&\omit &&\omit &\cr
&$\Omega ^1_{\ZZ}$ &&$1$ &&log derivative of map&&yes &\cr
height3pt &\omit &&\omit &&\omit &&\omit  &\cr
\noalign{\hrule}
}
\hrule}
\tag{6.2} $$
\noindent
 An object in the first line is a map $g\:M\to \TT$.  The corresponding field
strength~$d\log g$ is a global closed 1-form on~$M$ whose periods are integer
multiples of~$2\pi i$.  Note that \thetag{6.2}~is the special case of the
last two lines of~\thetag{6.1} when the circle bundle with connection is
trivial and trivialized.

The $B$-field on spacetime~$Y$ is an element of~$\Theta ^2(Y)$.  In other
words, it is the $p=2$~analog of the first line of~\thetag{6.1}
and~\thetag{6.2}.  Topologically, it is classified by a characteristic class
$\zeta_B\in H^3(Y;\ZZ)$, analogous to the Chern class of a circle bundle in
integral~$H^2$.  Geometrically, we describe~$B$ in terms of an open
cover~$\{U_i\}$ of~$Y$.  On each open set~$U_i$ there is a 2-form $B_i\in
\Omega ^2(Y_i)$, but on overlaps $U_{ij}=U_i\cap U_j$ they do not necessarily
agree.  Rather, there is a 1-form $\alpha _{ij}\in \Omega ^1(U_{ij})$ such
that
  $$ B_i-B_j=d\alpha _{ij}\qquad \text{on~$U_{ij}$}. \tag{6.3} $$
Similarly, on the triple intersection $U_{ijk}=U_i\cap U_j\cap U_k$ there is
given a circle-valued function $g_{ijk}\:U_{ijk}\to\TT$ such that
  $$ d\log g_{ijk} = \exp\bigl[ \sqrt{-1} (\alpha _{jk} - \alpha _{ik} +
     \alpha _{ij})\bigr]\qquad \text{on $U_{ijk}$}. \tag{6.4} $$ 
The $g_{ijk}$~satisfy the cocycle relation
  $$ g\mstrut _{jk\ell }\; g_{ik\ell }\inv\; g\mstrut _{ij\ell }\; g
     _{ijk}\inv =1\qquad \text{on $U_{ijk\ell }$}. \tag{6.5} $$
The $B$-field is the triple $B=\{B_i,\alpha _{ij},g_{ijk}\}$, which may be
conveniently placed in a double complex:
  $$  \vcenter{\offinterlineskip
      \halign{\hfil#\hfil\bigstrut\quad&\vrule\quad \hfil #\hfil\quad& \hfil
     #\hfil\quad& \hfil #\hfil\quad&\hfil #\hfil\quad& \hfil #\hfil\quad&\hfil
     #\hfil\quad& \hfil #\hfil \cr
      2&$B_i $&$\to$&\boxed{0}\cr
      \omit&\tinystrut\cr
      \omit&&&$\uparrow$\cr
      1&&&$\alpha _{ij}$&$\to$&\boxed{0}\cr
      \omit&\tinystrut\cr
      \omit&&&&&$\uparrow$\cr
      0&&&&&$g_{ijk}$&$\to$&\boxed{0}\cr
      \omit&\tinystrut\cr
      \noalign{\hrule}
      &$U_i$&\omit&\omit$U_{ij}$&&$U_{ijk}$&&$U_{ijk\ell }$\cr
      }}
      \tag{6.6} $$
Note $dB_i=dB_j=H$ is a global 3-form which is closed with $2\pi i\ZZ$
periods; it is the field strength of the $B$-field.  The boxes around the
three zeros in~\thetag{6.6} are shorthand for equations
\thetag{6.3}--\thetag{6.5}.   The horizontal arrows denote the \v Cech
differential and the vertical arrows denote plus or minus the de Rham
differential~$d$.

In the bosonic string with a $D$-brane, the $A$-field on the brane $\iota
\:Q\hookrightarrow Y$ is an element of~$\Gamma ^1(Q)$, that is, a
generalization of the third line of~\thetag{6.1} to~$p=1$.  However, it is
related to the restriction of the $B$-field to the brane in a specific way.
Namely, in terms of the restriction of the covering~$\{U_i\}$ to~$Q$, we
write $A=\{A_i,h_{ij}\}$ in the diagram:
  $$  \vcenter{\offinterlineskip
      \halign{\hfil#\hfil\bigstrut\quad&\vrule\quad \hfil #\hfil\quad& \hfil
     #\hfil\quad& \hfil #\hfil\quad&\hfil #\hfil\quad& \hfil #\hfil\quad&\hfil
     #\hfil\quad& \hfil #\hfil \cr
      2&$\iota ^*B_i $\cr
      \omit&\tinystrut\cr
      \omit&&&\phantom{$\uparrow$}\cr
      1&$A _{i}$&$\to$&\boxed{\hbox{$\iota ^*\alpha _{ij}$}}\cr
      \omit&\tinystrut\cr
      \omit&&&$\uparrow$\cr
      0&&&$h_{ij}$&$\to$&\boxed{\hbox{$\iota ^*g_{ijk}$}}\cr
      \omit&\tinystrut\cr
      \noalign{\hrule}
      &$\iota ^*U_i$&\omit&\omit$\iota ^*U_{ij}$&&$\iota ^*U_{ijk}$\cr
      }}
       \tag{6.7} $$
We impose equations at the two boxes.  Specifically,
  $$ \alignedat2
      A_j-A_i - d\log{h_{ij}} &= \iota ^*\alpha _{ij}&&\qquad \text{on $\iota
     ^*U_{ij}$}, \\
      h\mstrut _{jk}\;h_{ik}\inv \;h\mstrut _{ij}&=\iota ^*g_{ijk}&&\qquad
     \text{on $\iota ^*U_{ijk}$}.\endaligned \tag{6.8} $$
We make two remarks about these equations.  First, if $B=0$ so that the right
hand sides of~\thetag{6.8} vanish, then $\{h_{ij}\}$~are the transition
functions of a circle bundle and $\{A_i\}$~patch to a connection on it.  In
other words, if $B=0$ then the $A$-field is a connection on a circle
bundle---an element of~$\Theta ^1(Q)$ as in the first line of~\thetag{6.1}.
This is the usual interpretation of the gauge field on the brane.  Second,
for general $B$-fields we can say that $A$~is a ``trivialization'' of the
restriction~$\iota ^*B$, where the precise meaning of ``trivialization''
is~\thetag{6.8}.  By analogy consider the case one degree lower, where $B$~is
a circle bundle with connection and $A$~a trivialization.  There are two
possible meanings to ``trivialization'' in this context: {\it topological\/}
and {\it geometric\/}.  A topological trivialization corresponds to the
second equation of~\thetag{6.8} only.  A geometric trivialization corresponds
to imposing three equations---the equations of~\thetag{6.8} plus the equation
$dA_i=\iota ^*B_i$.  The intermediate case of two equations which we use has
no analog one degree lower.  Note that in our case the difference $dA_i -
\iota ^*B_i$ is a global 2-form which is not necessarily closed.  It is the
analog of the covariant derivative in the fourth line of~\thetag{6.1}.  For
$B=0$~it is the curvature of the abelian gauge field~$A$.

Now we make contact with the discussion in~\S{1}.  For the bosonic string the
term~\thetag{1.10} in the action is a well-defined number for each field
configuration.  One degree down, where $B$~is a connection on a circle bundle
and $A$~a trivialization of~$\iota ^*B$, this term corresponds to the
parallel transport along a path viewed as a number using the trivialization
on the boundary.  Hence this part of the action is well-defined as a
function---not a section of a line bundle---so does not contribute to the
anomaly.

In the superstring what was explained in~\S{1} is that since the first factor
in~\thetag{1.11} is a section of a nontrivial line bundle (over the space of
metrics on~$\Sigma $ and maps $\Sigma \to Y$), we must modify the global
interpretation of the $A$-field in order that the product of the last two
factors be a section of the inverse line bundle with connection.  In our
present context we describe the modification as follows.  First, the second
Stiefel-Whitney class~$w_2(\nu )$ of the normal bundle~$\nu $ to~$Q$ in~$Y$
determines a {\it flat\/} $B_{w_2(\nu )}\in \Theta ^2(Q)$, defined
in~\thetag{1.13}.  In terms of the open covering~$\{\iota ^*U_i\}$ it can be
written as
  $$  \vcenter{\offinterlineskip
      \halign{\hfil#\hfil\bigstrut\quad&\vrule\quad \hfil #\hfil\quad& \hfil
     #\hfil\quad& \hfil #\hfil\quad&\hfil #\hfil\quad& \hfil #\hfil\quad&\hfil
     #\hfil\quad& \hfil #\hfil \cr
      3&\boxed{0}\cr
      \omit&\tinystrut\cr
      \omit&$\uparrow$\cr
      2&0 &$\to$&\boxed{0}\cr
      \omit&\tinystrut\cr
      \omit&&&$\uparrow$\cr
      1&&&$0$&$\to$&\boxed{0}\cr
      \omit&\tinystrut\cr
      \omit&&&&&$\uparrow$\cr
      0&&&&&$w_{ijk}$&$\to$&\boxed{0}\cr
      \omit&\tinystrut\cr
      \noalign{\hrule}
      &$\iota ^*U_i$&\omit&\omit$\iota ^*U_{ij}$&&$\iota
      ^*U_{ijk}$&&$\iota ^*U_{ijk\ell }$\cr
      }}
       $$
where $\{w_{ijk}=\pm1\}$ is a \v Cech cocycle for~$w_2(\nu )$.  Then the
$A$-field is an isomorphism
  $$ A\:B_{w_2(\nu )} \longrightarrow \iota ^*B,  $$
where ``isomorphism'' is understood in the sense of equations~\thetag{6.8},
as explained above.  In other words, $A=\{A_i,h_{ij}\}$ fits into a slight
modification of~\thetag{6.7}:
  $$  \vcenter{\offinterlineskip
      \halign{\hfil#\hfil\bigstrut\quad&\vrule\quad \hfil #\hfil\quad& \hfil
     #\hfil\quad& \hfil #\hfil\quad&\hfil #\hfil\quad& \hfil #\hfil\quad&\hfil
     #\hfil\quad& \hfil #\hfil \cr
      2&$\iota ^*B_i $\cr
      \omit&\tinystrut\cr
      \omit&&&\phantom{$\uparrow$}\cr
      1&$A _{i}$&$\to$&\boxed{\hbox{$\iota ^*\alpha _{ij}$}}\cr
      \omit&\tinystrut\cr
      \omit&&&$\uparrow$\cr
      0&&&$h_{ij}$&$\to$&\boxed{\hbox{$\iota ^*g_{ijk}/w_{ijk}$}}\cr
      \omit&\tinystrut\cr
      \noalign{\hrule}
      &$\iota ^*U_i$&\omit&\omit$\iota ^*U_{ij}$&&$\iota ^*U_{ijk}$\cr
      }}
       $$
With this $A$-field \thetag{1.10}~is no longer a function, but rather a
section of a flat line bundle with connection over the space of parameters.
The holonomy of that bundle is computed by integrating $-w_2(\nu )=w_2(\nu )$
over the boundary of the surface mapping in, and this precisely cancels the
line bundle with connection from the fermionic determinants (as computed in
\theprotag{5.6} {Theorem}).  Thus the product of terms in~\thetag{1.11} is a
section of a trivializable line bundle with connection: there is no anomaly.

The {\it existence\/} of an isomorphism in the sense of
equations~\thetag{6.8} is equivalent to the existence of a topological
isomorphism, which only exists if
  $$ \iota ^*\zeta_B = W_3(\nu ), \tag{6.9} $$
where $\zeta_B\in H^3(Y;\ZZ)$ is the topological characteristic class of the
$B$-field and $W_3(\nu )$~is the third Stiefel-Whitney class of the normal
bundle.  Equation~\thetag{6.9}, which is the same as~\thetag{1.12}, is a
topological restriction on branes which may occur in a spacetime~$Y$ with
given $B$-field.  It was first discovered in a nonperturbative setting
in~\cite{W3}; the present paper derives~\thetag{6.9} from the perturbative
string.

Note in particular that if $B=0$, then \thetag{6.9}~asserts that $\nu
$~admits a $\spinc$~structure.  In that case the $A$-field, which
trivializes~$B_{w_2(\nu )}$ in the sense we described here, is a
$\spinc$~connection.  (The reader should relate our \v Cech description with
other definitions of $\spinc$~connections.)

 \bigskip\noindent{\it Dirac Operators with Local Boundary
Conditions\footnote{We are indebted to Xianzhe Dai for discussions about the
issues treated here.}}
\smallskip \nobreak
 Immediately following the first proof of the index theorem on closed
manifolds, Atiyah, Bott, and Singer~\cite{AB} proved a topological index
theorem for general elliptic operators with {\it local\/} boundary
conditions.  They observed that there is a topological obstruction to the
existence of local boundary conditions, and that local boundary conditions,
when they exist, lift the symbol class of the operator in $K$-theory to a
relative class.  It is in terms of this lifted class that one obtains a
topological index formula.  Later, Atiyah, Patodi, and Singer~\cite{APS1}
introduced {\it global\/} boundary conditions for first-order Dirac
operators, and these always exist.  Perhaps for that reason it is global,
rather than local, boundary conditions which appear in most of the index
theory literature.  In this subsection we generalize the local boundary
conditions discussed in~\S{5} to Dirac operators in arbitrary even
dimensions, and then we indicate how geometric aspects of index theory may be
treated by doubling.  (Doubling is a common technique in the theory of
elliptic boundary-value problems in flat space as well, and it is also the
main technique in the Atiyah-Bott paper.)  We only indicate a rough outline
of the arguments; we have not carried through the details.  For simplicity we
deal only with complex Dirac operators, though the discussion applies to real
Dirac operators as well.

As an aside, we remark that generally in physics boundary conditions
play at least two different roles.  An object (such as a $D$-brane or
the Earth's ocean) may have a boundary, at which one imposes local boundary
conditions.  The normal vector to such a boundary is ordinarily spacelike.
On the other hand, an important physical and mathematical technique
is to ``cut'' on a spacelike surface to reveal a quantum state.
The normal vector to such a boundary is generally timelike if we work
with Lorentz signature, and otherwise spacelike.  On such a cut, one
uses global boundary conditions, similar to those used in index theory
in factorization theorems, such as the one leading to \thetag{3.5}.

Consider an even-dimensional $\spinc$~manifold~$X$ with boundary.  Let
$S^{\pm}$~denote the spinor bundles on~$X$.  Let~$\zeta $ be the unit outward
normal vector field at the boundary.  The Clifford multiplication
  $$ c(\zeta )\:S^+\res{\bX}\longrightarrow S^-\res{\bX} \tag{6.10} $$
is an isomorphism.  Suppose $E^{\pm}\to X$ are complex vector bundles (with
connection) and we are given an isomorphism
  $$ \tau \:E^+\res{\bX}\longrightarrow E^-\res{\bX}  $$
at the boundary.  Then the Dirac operator
  $$ D^X\:S^+\otimes E^+ \;\oplus \; S^-\otimes E^-\longrightarrow S^-\otimes
     E^+ \;\oplus \;S^+\otimes E^- \tag{6.11} $$
admits the local boundary condition
  $$ \bigl(c(\zeta )\otimes \tau  \bigr)\bigl(\psi ^+\res{\bX} \bigr) = \psi
     ^-\res{\bX} \tag{6.12} $$
on a pair~$(\psi ^+,\psi ^-)$ of sections of~$(S^+\otimes E^+,S^-\otimes
E^-)$.  The boundary-value problems considered in section~5 are special
cases.  Note that different spin (or $\text{spin}^c$) structures on~$S^+,S^-$
may be accommodated by tensoring~$E^-$ with a real (or complex) line bundle.
We propose that this class of Dirac operators has a good geometric index
theory, analogous to that of Dirac operators on closed manifolds.

It is not hard to check that \thetag{6.12}~defines {\it elliptic\/} boundary
conditions (as described in~\cite{AB}, for example).  Therefore, the Dirac
operator enjoys the same basic analytic properties as those of a Dirac
operator on a closed manifold:  the spectrum is discrete, there is a
meromorphic $\zeta $-function, etc.  There is no problem carrying out
geometric constructions, such as the determinant line bundle with metric and
connection, in families of such operators.  From such constructions one will
obtain formulas for geometric invariants, such as curvature and holonomy of
the determinant line bundle, directly on~$X$.  In the remainder of this
section we indicate how to reduce these constructions to the closed case by
doubling.

 \midinsert
 \bigskip
 \centerline{
 \epsfxsize=300pt
 \epsffile{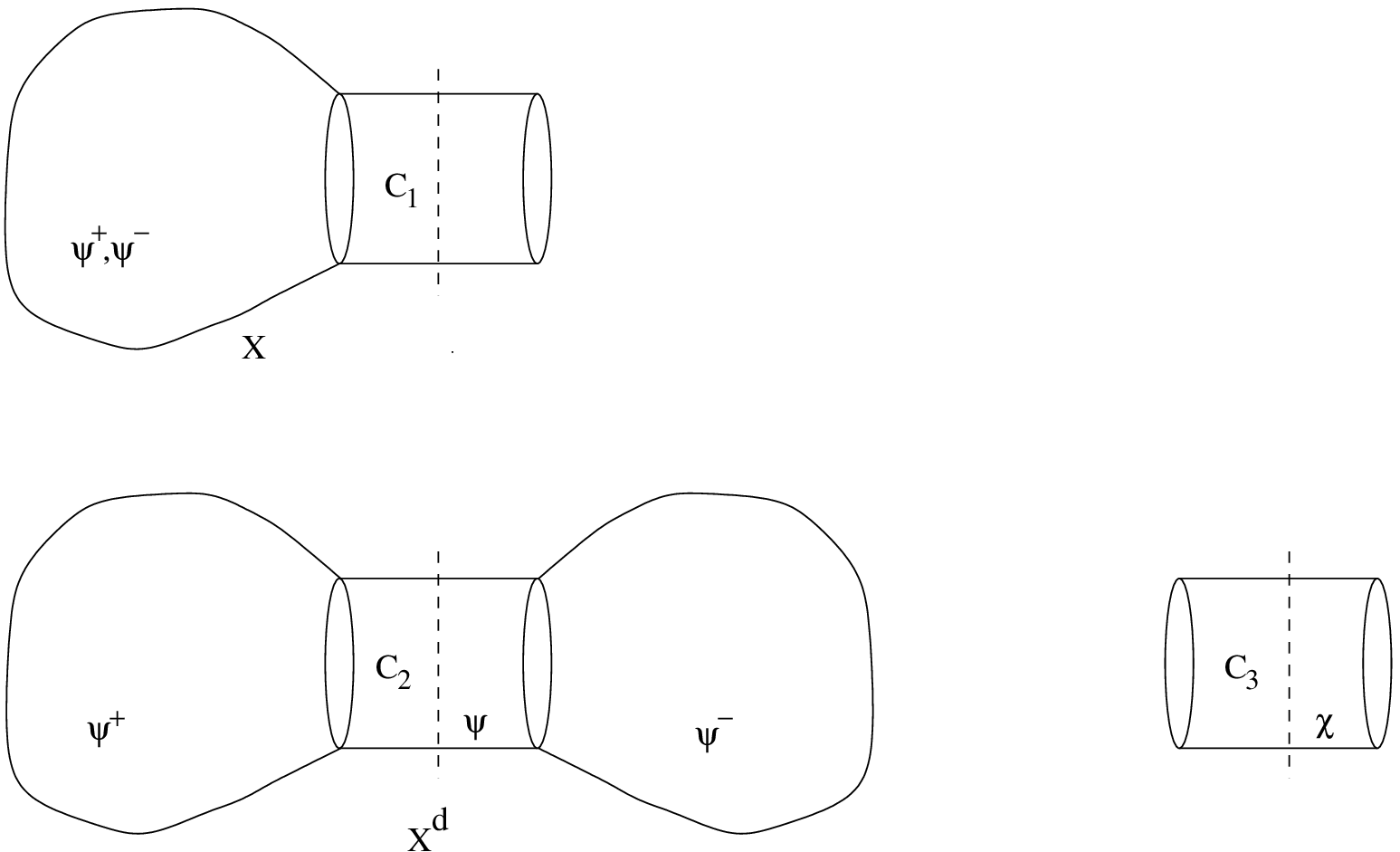}}
 \nobreak
 \botcaption{Figure~1: Gluing spinor fields}
 \endcaption
 \bigskip
 \endinsert

 \midinsert
 \bigskip
 \centerline{
  \epsfxsize=150pt
 \epsffile{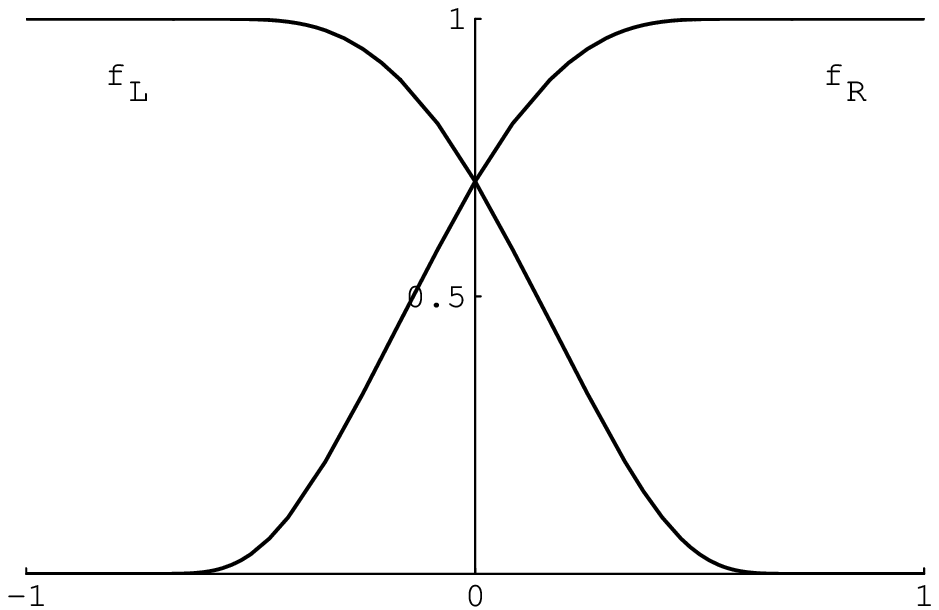}}
 \nobreak
 \botcaption{Figure~2: Cutoff functions}
 \endcaption
 \bigskip
 \endinsert

We restrict to manifolds~$X$ whose Riemannian metric is a product near the
boundary.  Let $\Xd$~denote the smooth closed Riemannian manifold obtained by
gluing~$X$ to~$-X$ along~$\bX$.  We glue $S^+\to X$ and~$S^-\to -X$
using~\thetag{6.10} to obtain the plus spinor bundle $S^+_{\Xd}\to \Xd$, and
similarly glue $E^+\to X$ to $E^-\to -X$ to form a complex vector bundle
$E\to \Xd$ (with connection).  Our goal is to precisely relate the
boundary-value problem~\thetag{6.11}, \thetag{6.12} with the Dirac operator
$D^+_{\Xd}(E)$ on the double.  We adapt the gluing argument
of~\cite{DF,\S IV}, which we summarize in Figures~1 and~2.  Let $\Gamma
$~denote the space of smooth sections, regarded as an inner product space
using the $L^2$~metric.  The figures summarize a gluing map
  $$ U\:\Gamma _X(S^+\otimes E^+\;\oplus \; S^-\otimes E^-) \longrightarrow
     \Gamma _{\Xd}(S^+\otimes E) \;\oplus \; \Gamma _{C_3}(S^+\otimes E),
      $$
which for appropriate cutoff functions is an isometry.  The spinor
fields~$(\psi ^+,\psi ^-)$ in the domain satisfy the boundary
condition~\thetag{6.12}, and the spinor field~$\chi $ on~$C_3$ satisfies
  $$ \chi (t=1) \;=\; -\chi (t=-1),  $$
where $t$~is the axial coordinate on the cylinder.  More explicitly, the
gluing map on cylinders is
  $$ \pmatrix \psi (t)\\\chi (t)  \endpmatrix = \pmatrix
     f_L(t)&f_R(t)\\-f_R(t)&f_L(t)  \endpmatrix \pmatrix \psi ^+(t)\\\psi
     ^-(-t)  \endpmatrix,\qquad \text{$\psi ^+,\psi ^-$ on~$C_1$}, \quad
     \text{$\psi$ on~$C_2$}, \quad \text{$\chi $ on~$C_3$}.   $$
The cutoff functions $f_L,f_R\:[-1,1]\to[0,1]$ satisfy
  $$ \alignedat2
      f_L\bigl([-1,-1/2] \bigr) &= f_R\bigl([1/2,1] \bigr) = 1&\qquad \qquad
      f_L\bigl([1/2,1] \bigr) &= f_R\bigl([-1,-1/2] \bigr) = 0\\
      f_L^2 + f_R^2 &= 1&\qquad \qquad
      f_L(-x) &= f_R(x).\endalignedat $$
The same gluing works in a family of operators; the result is an isometry of
infinite dimensional vector bundles of spinor fields.

The claim is that the determinant line bundle for the Dirac operator on~$C_3$
is canonically trivial, and so, after some more argument, in such a family
the determinant line bundle (with its metric and connection) for the
boundary-value problem on~$X$ is canonically isomorphic to the determinant
line bundle for the closed manifold~$X^d$.

So as not to rely on the details of this argument, which we may present
elsewhere, in this paper we adapted the practical point of view of {\it
defining\/} the metric and connection directly from the Dirac operator on the
double.


\Refs\tenpoint
\vskip-6pt

\ref
\key A
\by J. F. Adams \paper Vector-fields on spheres\jour Bull. Amer. Math. Soc \vol
68 \yr 1962 \pages 39--41
\endref

\ref
\key AB
\by M. F. Atiyah, R. Bott
\paper The index problem for manifolds with boundary
\inbook  Differential Analysis
\bookinfo Bombay Colloq., 1964
\pages 175--186
\publ Oxford Univ. Press
\publaddr London
\yr 1964
\endref

\ref
\key APS1
\by M. F. Atiyah, V. K. Patodi, I. M. Singer
\paper Spectral asymmetry and Riemannian geometry. I
\jour Math. Proc. Cambridge Philos. Soc. \vol 77 \yr 1975 \pages 43--69
\endref

\ref
\key APS2
\by M. F. Atiyah, V. K. Patodi, I. M. Singer
\paper Spectral asymmetry and Riemannian geometry. III
\jour Math. Proc. Cambridge Philos. Soc. \vol 79
\yr 1976 \pages 71--99
\endref

\ref
\key AS1
\by M. F. Atiyah, I. M. Singer \paper The index of elliptic operators III\jour
Ann. of Math. \vol 87 \yr 1968 \pages 546--604
\endref

\ref
\key AS2
\by M. F. Atiyah, I. M. Singer \paper The index of elliptic operators V\jour
Ann. of Math. \vol 93 \yr 1971 \pages 139--149
\endref

\ref
\key B
\by J.-L. Brylinski
\book  Loop spaces, characteristic classes and geometric quantization
\bookinfo Progr. Math., 107
\publ Birkh\"auser
\publaddr Boston, MA
\yr 1993
\endref

\ref
\key BF
\by J. M. Bismut, D. S. Freed \paper The analysis of elliptic
families I: Metrics and connections on determinant bundles \jour Commun. Math.
Phys. \vol 106 \pages 159--176 \yr 1986
\moreref \paper II: Dirac operators, eta invariants, and the holonomy theorem
of Witten \jour Commun. Math. Phys. \vol 107 \yr 1986 \pages 103--163
\endref

\ref
\key BW
\by B. Booss-Bavnbek, K. P. Wojciechowski \book Elliptic Boundary Problems
for Dirac Operators \publ Birkh\"auser \publaddr Boston \yr 1993
\endref

\ref
\key BH
\by A. Borel, F. Hirzebruch \paper Characteristic classes and homogeneous
spaces. II \jour Amer. J. Math. \vol 81 \yr 1959 \pages 315--382
\endref

\ref
\key CY
\by Y.-K. E. Cheung and Z. Yin,
\paper Anomalies, Branes, and Currents
\jour Nucl. Phys. \vol B517 \yr 1998 \pages 185-196
\finalinfo {\tt hep-th/9803931}
\endref


\ref
\key DF
\by X. Dai, D. S. Freed
\paper $\eta $-invariants and determinant lines
\jour J. Math. Phys.
\vol 35
\yr 1994
\pages 5155--5194
\endref

\ref
\key DZ
\by X. Dai, W. Zhang
\paper Splitting of the family index
\jour Commun. Math. Phys.
\vol 182
\yr 1996
\pages 303--317
\endref

\ref
\key DeF
\by{P. Deligne, D. S. Freed}
\paper{Classical field theory}
\inbook{Quantum Fields and Strings: A Course for Mathematicians}
 \eds{P. Deligne, P. Etingof, D. S. Freed, L. C. Jeffrey, D. Kazhdan,
J. W. Morgan, D. R. Morrison, E. Witten}
 \publ{American Mathematical Society}
 \yr{1999}
 \publaddr{Providence, RI}
 \bookinfo{Volume~1}
\pages{137--225}
\endref


\ref
\key F1
\by D. S. Freed \paper On determinant line bundles \inbook Mathematical Aspects
of String Theory \bookinfo ed. S. T. Yau \publ World Scientific Publishing \yr
1987
\endref

\ref
\key F2
\by  D. S. Freed \paper Determinants, torsion, and strings
\jour Commun. Math. Phys \vol 107 \yr 1986 \pages 483--513
\endref

\ref
\key F3
\by D. S. Freed
\paper Two index theorems in odd dimensions
\jour Commun. Anal. Geom.
\vol 6
\yr 1998
\pages 317--329
\finalinfo {\tt dg-ga/9601005}
\endref

\ref
\key FM
\by D. S. Freed, J. W. Morgan
\paper Appendix to \cite{F2}
\jour Commun. Math. Phys \vol 107 \yr 1986 \pages 510--513
\endref

\ref
\key GHM
\by M. Green, J. A. Harvey, and G. Moore
\paper $I$-Brane Inflow And Anomalous Couplings On $D$-Branes
\jour Class. Quant. Grav. \vol 14 \yr 1997 \pages 47-52
\finalinfo {\tt hep-th/9605033}
\endref

\ref
\key H
\by P. Horava
\paper Type IIA $D$-Branes, $K$ Theory, And Matrix Theory
\jour Adv. Theor. Math. Phys. 
\vol 2 
\yr 1999 
\pages 1373--1404
\finalinfo {\tt hep-th/9812135}
\endref


\ref
\key MM
\by G. Moore and R. Minasian
\paper K Theory And Ramond-Ramond Charge
\jour JHEP  9711:002, 1997
\finalinfo {\tt hep-th/9710230}
\endref

\ref
\key N
\by{Nicolaescu, L. I.}
\paper{Generalized symplectic geometries and the index of families of
             elliptic problems}
\jour{Mem. Amer. Math. Soc.}
\vol{128}
\issue{609}
\yr{1997}
\endref

\ref
\key P
\by P. Piazza
\paper Determinant bundles, manifolds with boundary and surgery
\jour Commun. Math. Phys.
\vol 178
\yr 1996
\pages 597--626
\moreref
\paper II. Spectral sections and surgery rules for
anomalies
\jour Commun. Math. Phys.
\vol 193
\yr 1998
\pages 105--124
\endref

\ref
\key S
\by A. Sen
\paper Tachyon Condensation On The Brane Antibrane System
\jour JHEP 9808:012, 1998
\finalinfo {\tt hep-th/9805170}
\endref



\ref
\key W1
\by E. Witten \paper Global gravitational anomalies \jour Commun. Math. Phys.
\vol 100 \yr 1985 \pages 197--229
\endref

\ref
\key W2
\by E. Witten \paper Global anomalies in string theory \inbook Anomalies,
Geometry, and Topology \bookinfo edited by A. White \publ World Scientific \yr
1985 \pages 61--99
\endref

\ref
\key W3
\by E. Witten \paper Baryons And Branes In Anti-de Sitter Space
\jour JHEP 9807:006, 1998
\finalinfo {\tt hep-th/9805112}
\endref

\ref\
\key W4
\by E. Witten \paper $D$-Branes And $K$-Theory
\jour JHEP 9812:025, 1998
\finalinfo{\tt hep-th/9810188}
\endref

\endRefs
\enddocument